\documentclass[twocolumn,tighten,times]{aastex61}
\usepackage{bm}
\usepackage{amsmath}
\usepackage{color}
\usepackage{enumitem}

\shorttitle{The $\rho$ Oph-A Core}
\shortauthors{Kwon et al.}

\submitjournal{To appear in the Astrophysical Journal}

\begin{document}

\fontsize{10}{10.6}\selectfont

\title{A First Look at BISTRO Observations of The $\rho$ Oph-A core}
       
\AuthorCallLimit=130

\correspondingauthor{Jungmi Kwon}
\email{jungmi.kwon@ir.isas.jaxa.jp}

\author[0000-0003-2815-7774]{Jungmi Kwon}
\affil{Institute of Space and Astronautical Science, Japan Aerospace Exploration Agency, 3-1-1 Yoshinodai, Chuo-ku, Sagamihara, Kanagawa 252-5210, Japan}

\author{Yasuo Doi}
\affiliation{Department of Earth Science and Astronomy, Graduate School of Arts and Sciences, The University of Tokyo,　3-8-1 Komaba, Meguro, Tokyo 153-8902, Japan}

\author{Motohide Tamura}
\affiliation{Department of Astronomy, Graduate School of Science, The University of Tokyo, 7-3-1 Hongo, Bunkyo-ku, Tokyo 113-0033, Japan}
\affiliation{Astrobiology Center, National Institutes of Natural Sciences, 2-21-1 Osawa, Mitaka, Tokyo 181-8588, Japan}
\affiliation{National Astronomical Observatory of Japan, National Institutes of Natural Sciences, Osawa, Mitaka, Tokyo 181-8588, Japan}

\author{Masafumi Matsumura}
\affiliation{Kagawa University, Saiwai-cho 1-1, Takamatsu, Kagawa, 760-8522, Japan}

\author{Kate Pattle}
\affiliation{Institute of Astronomy and Department of Physics, National Tsing Hua University, Hsinchu 30013, Taiwan}
\affiliation{Jeremiah Horrocks Institute, University of Central Lancashire, Preston PR1 2HE, UK}
\affiliation{National Astronomical Observatory of Japan, National Institutes of Natural Sciences, Osawa, Mitaka, Tokyo 181-8588, Japan}

\author{David Berry}
\affiliation{East Asian Observatory, 660 N. A`oh\={o}k\={u} Place, University Park, Hilo, HI 96720, USA}

\author{Sarah Sadavoy} 
\affiliation{Harvard-Smithsonian Center for Astrophysics, 60 Garden Street, Cambridge, MA 02138, USA}

\author{Brenda C. Matthews}
\affiliation{NRC Herzberg Astronomy and Astrophysics, 5071 West Saanich Road, Victoria, BC V9E 2E7, Canada}
\affiliation{Department of Physics and Astronomy, University of Victoria, Victoria, BC V8P 1A1, Canada}

\author{Derek Ward-Thompson}
\affiliation{Jeremiah Horrocks Institute, University of Central Lancashire, Preston PR1 2HE, UK}

\author{Tetsuo Hasegawa} 
\affiliation{National Astronomical Observatory of Japan, National Institutes of Natural Sciences, Osawa, Mitaka, Tokyo 181-8588, Japan}

% ...

\begin{abstract}
We present 850 $\mu$m imaging polarimetry data of
the $\rho$ Oph-A core 
taken with the Submillimeter Common-User Bolometer Array-2 (SCUBA-2) and its polarimeter (POL-2),
as part of our ongoing survey project, BISTRO ($\bm B$-fields In STar forming RegiOns).
The polarization vectors are used to identify the orientation of the magnetic field projected on the plane of the sky at a resolution of 0.01 pc.  We identify 10 subregions with distinct polarization fractions and angles in the 0.2 pc $\rho$ Oph A core;
some of them can be part of a coherent magnetic field structure in the $\rho$ Oph region.
The results are consistent with previous observations of the brightest regions of $\rho$ Oph-A,
where the degrees of polarization are at a level of a few percents,
but our data reveal for the first time the magnetic field structures 
in the fainter regions surrounding the core where the degree of polarization is much  higher ($> 5 \%$). 
A comparison with previous near-infrared polarimetric data shows that there are several magnetic field components 
which are consistent at near-infrared and submillimeter wavelengths. 
Using the Davis-Chandrasekhar-Fermi method, we also derive magnetic field strengths in several sub-core regions,
which range from approximately 0.2 to 5 mG.
We also 
find a correlation between the magnetic field
orientations projected on the sky with the core centroid velocity components.
\end{abstract}


\section{INTRODUCTION} \label{sec:intro}

Stars form in dense and cold molecular clouds and
it has long been considered that magnetic fields may play significant roles in various stages of star formation 
(e.g., \citealp{1987ARA&A..25...23S, 2007ARA&A..45..339B, 2007ARA&A..45..565M, 2013ASPC..476...95A}).
Near-infrared linear polarimetry is one of the traditional methods of tracing magnetic field structure in order to measure the magnetic fields 
in denser regions than those traced by optical polarimetry, which are directly related to the star formation process
(e.g., \citealp{1951ApJ...114..206D, 2007JQSRT.106..225L}). 
The magnetic field has been successfully traced in dense regions of several molecular clouds
(e.g., \citealp{1979AJ.....84..199W, 1987MNRAS.224..413T, 1988MNRAS.231..445T, 2007PASJ...59..467T, 2010ApJ...708..758K, 2011ApJ...741...35K, 2015ApJS..220...17K, 2014ApJ...793..126C, 2014ApJ...783....1S, 2017ApJ...842...66W}).
Polarization at near-infrared wavelengths, however, relies on measurements of dust extinction from background stars 
and as such cannot trace well magnetic fields in denser substructures like filaments and cores within clouds.  
As these structures are directly linked to star formation, it is vital to measure their magnetic fields.  
Observations of dust polarization from thermal emission at far-infrared and (sub)millimeter wavelengths 
can trace these high column densities and probe how the magnetic field influences the star formation process
(e.g., \citealp{1999ApJ...525..832T, 2015MNRAS.450.1094P, 2017PKAS...32..117W}; see also \citealt{2016A&A...596A..93S}).

The $\rho$ Ophiuchi (hereafter $\rho$ Oph) dark cloud complex
is one of the closest star-forming regions at a distance of approximately 120--165 pc 
(e.g., \citealp{1981A&A....99..346C, 1989A&A...216...44D, 1998A&A...338..897K, 2004AJ....127.1029R, 2008ApJ...675L..29L, 
2008A&A...480..785L, 2008AN....329...10M, 2008ApJ...679..512S, 2017ApJ...834..141O}). 
It has also been widely studied 
(see \citealp{2015ApJS..220...17K, 2008hsf2.book..351W}, for a reference summary).
It is a nearby region of clustered low- to intermediate-mass star formation (e.g., \citealp{2008hsf2.book..351W})
and is heavily influenced by the nearby Sco OB2 association 
(\citealp{1977AJ.....82..198V, 1989ApJ...338..902L, 1989ApJ...338..925L, 2015ApJS..220...17K}).
It was observed as part of the JCMT Gould Belt Legacy Survey \citep{2007PASP..119..855W}, 
the Herschel Gould Belt Survey \citep{2010A&A...518L.102A}, and the Spitzer Gould Belt Survey \citep{2009ApJS..181..321E}.
In the main body of $\rho$ Oph, 
detailed DCO$^+$ observations have identified several very dense, cold cores labeled A--F 
(\citealp{1986ApJ...306..142L, 1990ApJ...365..269L}), and
$\rho$ Oph-A appears to be the warmest among these cores \citep{1984A&A...141..127Z}.
The first submillimeter continuum observations of the $\rho$ Oph-A core region were obtained by \citet{1989MNRAS.241..119W}.
Many sub-cores in this region were identified (e.g., \citealt{2007A&A...472..519A, 1998A&A...336..150M}), 
which will be described in Section \ref{sec:discussion}. 
In this paper, we use the term ``core" for the $\rho$ Oph-A complex and the term ``sub-core" for the smaller condensations within it.

Here we present new observations of the $\rho$ Oph-A core in dust polarization from the James Clerk Maxwell Telescope (JCMT) 
as part of the $\bm B$-fields In STar forming RegiOns (BISTRO) survey \citep{2017ApJ...842...66W}.
The JCMT magnetic field survey of the Gould Belt clouds 
is a large-scale project, which aims to map
the submillimeter polarization of the dust thermal emission in the densest parts 
of all of the Gould Belt star forming regions.
The combination of SCUBA-2 (Submillimeter Common-User Bolometer Array-2; \citealp{2013MNRAS.430.2513H}) and its polarimeter POL-2 (Bastien et al, in prep.) enables deep submillimeter polarimetry 
and
is one of the most powerful instruments to reveal the magnetic field structure in star forming regions
thanks to its high sensitivity and high resolution (\citealt{2017ApJ...842...66W, 2017ApJ...846..122P}).

The paper is outlined as follows:
In Section \ref{sec:obs}, we describe the submillimeter observations, 
and the SCUBA-2/POL-2 data reduction is described in Section \ref{sec:dr}.
In Section \ref{sec:results}, we present the results of the submillimeter imaging polarimetry. 
In Section \ref{sec:discussion}, we discuss the magnetic field structure related to the star-forming activity in the $\rho$ Oph-A core region. 
A summary is given in Section \ref{sec:summary}.

\section{OBSERVATIONS} \label{sec:obs}

Continuum observations of $\rho$ Oph-A at 850 $\mu$m 
were made by inserting POL-2 into the optical path of SCUBA-2
between 2016 April 15 and 2016 April 24.
The region was observed in 20 sets of 41-minute observations
and among the 20 sets, 2 sets with bad quality data were excluded.
Note that the BISTRO time was allocated to take place during Band 2 weather ($0.05 < \tau_{225 \rm \, GHz} < 0.08$).
The observations were made using fully-sampled 12$\arcmin$ diameter circular regions with a resolution of 14$\farcs$1 
using a version of the SCUBA-2 DAISY mapping mode \citep{2013MNRAS.430.2513H} optimized for POL-2 observations.
The POL-2 DAISY scan pattern produces a central 3$\arcmin$ diameter region of approximately even coverage, 
with noise increasing to the edge of the map. 
The mode has a scan speed of 8$\arcsec$/s, with a half-waveplate rotation speed of 2 Hz \citep{2016SPIE.9914E..03F}. 
Continuum polarimetric observations were simultaneously taken at 450 $\mu$m with a resolution of 9.6$\arcsec$.
In this paper we discuss 850 $\mu$m data only.

\section{DATA REDUCTION} \label{sec:dr}

The 850 $\mu$m POL-2 data were reduced in a three-stage process using the $pol2map$ routine (the version updated on 2017 May 27) in \textsc{smurf} (\citealt{2005ASPC..343...71B, 2013MNRAS.430.2545C}), which we summarize here.  
POL-2 data reduction is described in detail by Bastien et al. (in prep.). 
See also \citealt{2017ApJ...842...66W} for a brief summary.

In the first stage, the raw bolometer timestreams for each observation are converted into separate Stokes $Q$, $U$, and $I$ timestreams using the process $calcqu$.  An initial Stokes $I$ map is created from the $I$ timestream from each observation using the iterative map-making routine $makemap$.  For each reduction, areas of astrophysical emission are defined using a signal-to-noise-based mask determined iteratively by $makemap$.  Areas outside this masked region are set to zero 
until the final iteration of $makemap$ (see \citealt{2015MNRAS.454.2557M} for a detailed description of the role of masking in SCUBA-2 data reduction).  Each map is compared to the first map in the sequence to determine a set of relative pointing corrections.  The individual $I$ maps are coadded to produce an initial $I$ map of the region.

In the second stage, an improved Stokes $I$ map is created from the $I$ timestreams of each observation using $makemap$.  
The initial $I$ map (described above) is used to generate a fixed signal-to-noise-based mask 
for all iterations of $makemap$.  The pointing corrections determined in Stage 1 are applied during the map-making process. 
In all cases, the polarized sky background is estimated by doing a
Principal Component Analysis (PCA) of the $I$, $Q$, and $U$ timestreams to identify components that are common to multiple bolometers. 
In the first stage, the 50 most correlated components are removed at each iteration. 
In the second stage 150 components are removed at each iteration, 
resulting in smaller changes in the map between iterations and lower noise in the final map.
All of the individual improved $I$ maps are co-added to form the final output $I$ map.

In the third stage, the Stokes $Q$ and $U$ maps, and the final vector catalogue, are created. Individual $Q$ and $U$ maps are reduced separately using $makemap$, and are created from the timestreams created in Stage 1, using the same mask based on the initial Stokes $I$ map as was used in Stage 2, and using the pointing offsets determined in Stage 1.  Correction for instrumental polarization is performed, based on the final output $I$ map.  The sets of individual $Q$ and $U$ maps are then coadded to create 
final $Q$ and $U$ maps.  
The final coadded Stokes $Q$, $U$ and $I$ maps are used to create an output vector catalogue, which includes 
the coordinates (J2000.0), 
values of Stokes parameters, 
degrees of polarization ($P \pm \delta P$), and
polarization position angles ($\theta \pm  \delta \theta$).
Therefore, it uses exactly the same map-making procedure to create all three maps -- Stokes $Q$, $U$, and $I$,
and spatial frequencies present in the three maps are all in common.

The output $Q$, $U$ and $I$ maps are gridded to 4$^{\prime\prime}$ pixels and are calibrated in mJy beam$^{-1}$
using a flux conversion factor (FCF) of 725 Jy\,pW$^{-1}$ (the standard SCUBA-2 850 $\mu$m FCF of 537 Jy\,pW$^{-1}$ multiplied by a factor of 1.35 to account for additional losses from POL-2; cf. \citealt{2013MNRAS.430.2534D, 2016SPIE.9914E..03F}). The output vectors are debiased using the mean of their $Q$ and $U$ variances to remove statistical biasing in regions of low signal-to-noise (see Equation \ref{eq:3} below).

The raw degree of polarization, $P'$, and the uncertainty in the degree of polarization, $\delta P$, 
can be calculated from the expressions:
\begin{equation}
P' = {\sqrt{Q^2 + U^2} \over I} \times 100 \%
\label{eq:1}
\end{equation}
and 
\begin{equation}
\delta P' = (P'I^2)^{-1} \sqrt{(Q^2 \delta Q^2+U^2 \delta U^2+(P')^4 I^2 \delta I^2)}
\label{eq:2}
\end{equation}
Note that in the pipeline software (without debiasing; see below), 
$P'I$ is first calculated from $Q$, $\delta Q$, $U$, and $\delta U$, 
then $\delta P'$ is calculated from $I$, $\delta I$, $Q$, $U$, and $\delta PI$. 
The expression
here is identical to the formula in the pipeline
but tries to show
the dependence on the errors of $I$, $Q$, and $U$.

As mentioned, a bias exists that tends to increase the polarization percentage value, 
even when Stokes $Q$ and $U$ are consistent with a value of zero 
because polarization percentage is forced to be positive
\citep{2006PASP..118.1340V}.
To mitigate this problem, 
approximate de-biased values are calculated in the pipeline,
assuming $\delta Q \sim \delta U$,
as follows:
\begin{equation}
PI = \sqrt{Q^2+U^2-0.5(\delta Q^2 + \delta U^2)}
\label{eq:3}
\end{equation}
and the degree of polarization $P$ is derived from the polarized intensity $PI$ as
\begin{equation}
P = PI / I.
\label{eq:4}
\end{equation}

The polarization position angles, $\theta$, and their errors, $\delta \theta$, can then be calculated by the following relations:
\begin{equation}
\theta = {1 \over 2} \, {\rm tan}^{-1} {U \over Q}
\label{eq:5}
\end{equation}
and 
\begin{equation}
\delta \theta = 0.5 \times \sqrt{ Q^2 \times \delta U^2 + U^2 \times \delta Q^2 } /( Q^2 + U^2 )\times 180\arcdeg/\pi
\label{eq:6}
\end{equation}

The data reduction process described above derives the Stokes $I$ map from the same POL-2 observations 
that are used to derive the Stokes $Q$ and $U$ maps. 
A consequence of this is that the FCF for the Stokes $I$, $Q$, and $U$ maps are then 
all equal and so cancel out when calculating the fractional polarization. 
As a result, the $I$ and the $Q$ and $U$ maps necessarily have exactly the same spatial scales.
Earlier versions of the POL-2 pipeline software derived the Stokes $I$ map from separate observations 
taken without POL-2 in the beam, resulting in the $I$ map having a different FCF to the Stokes $Q$ and $U$ maps 
because of the attenuation caused by POL-2 and differences in the map-making procedure (cf. \citealt{2016SPIE.9914E..03F}).

\section{Results} \label{sec:results}

\subsection{POL-2 Data Verification} \label{sec:verification}

The BISTRO survey has recently begun to systematically investigate magnetic field structures in the dense cores using measurements of polarized dust emission, which is one of the most effective ways of probing the magnetic fields of such cores.
Since POL-2 is newly commissioned, 
it is an important step to verify the consistency of our new data with those of previous studies.
Therefore, 
we compare the POL-2 observations of $\rho$ Oph-A with data from 
the SCUPOL polarimeter on the previous generation submillimeter bolometer array on the JCMT, SCUBA \citep{1999A&A...344..668G}.

\begin{figure*} 
\epsscale{1.}
\plotone{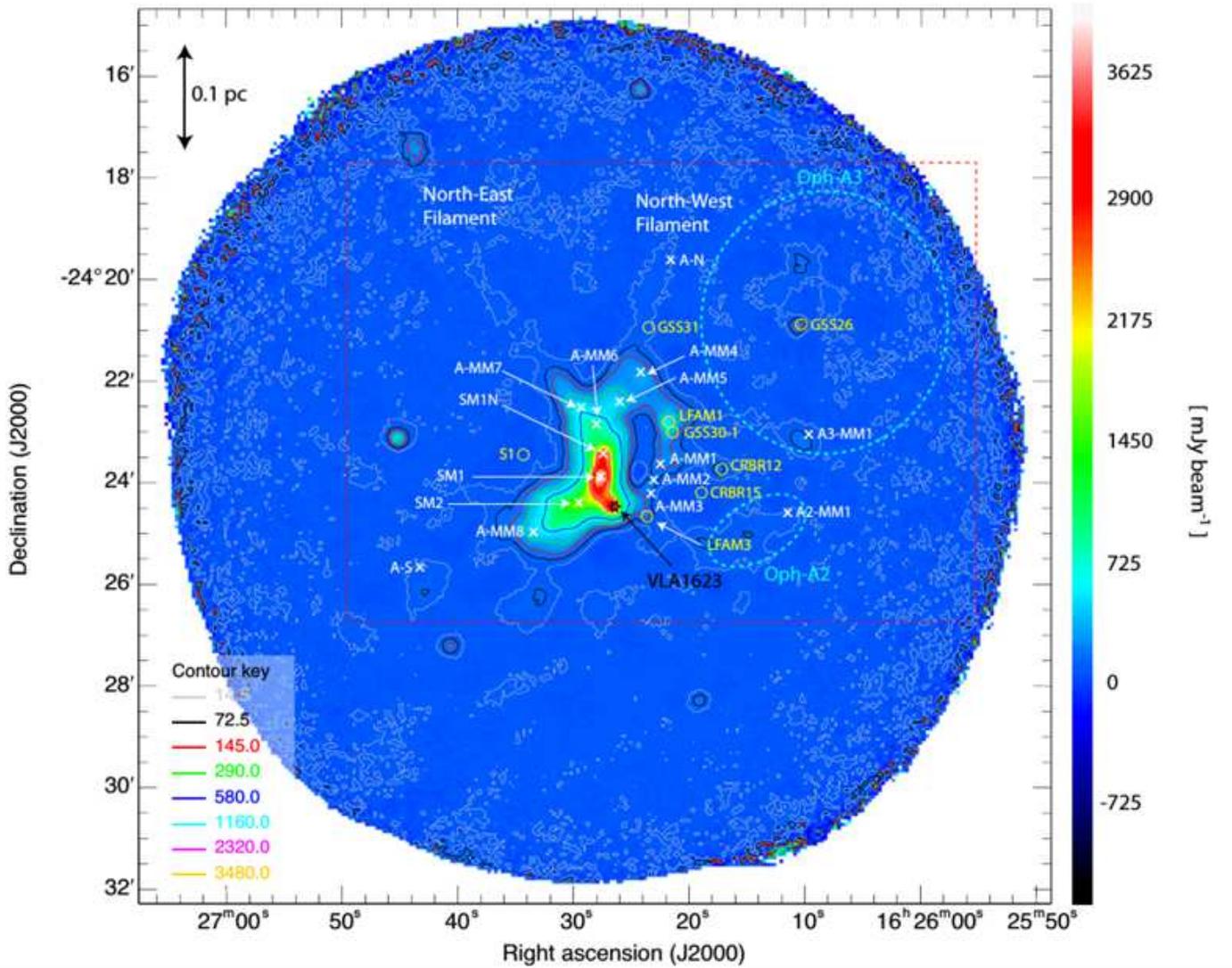}
\caption{
Stokes $I$ image (linear scale) of the $\rho$ Oph-A field centered on Oph-A1 
\citep{1998A&A...336..150M} obtained using JCMT with SCUBA-2/POL-2.
Notable sources and features in this region are labeled.
The spatial resolution is 14$\farcs$1 or approximately 0.01 pc assuming a distance to the Oph cloud of 140 pc.
The white crosses indicate the positions of the starless condensations
identified by \citet{1998A&A...336..150M} in the dust continuum at 1.2 mm 
(corresponding to the red dashed rectangle region in this figure),
the yellow circles indicate the position of young embedded stars,
and the black star indicates the position of VLA~1623.
The cyan dashed circles indicate Oph-A2 and Oph-A3, respectively, defined by \citet{1998A&A...336..150M}.
The contour levels are arbitrarily chosen to emphasize the Oph-A core
and their keys are shown in mJy beam$^{-1}$ unit.
}
\label{fig:OphA_stokesI}
\end{figure*}

Figure \ref{fig:OphA_stokesI} shows the 850 $\mu$m intensity map (Stokes $I$)
obtained using the JCMT with SCUBA-2/POL-2 
with well known submillimeter and infrared sources  labeled. The
Stokes $I$ image is consistent with previous deep submillimeter continuum images
(e.g., \citealt{2015MNRAS.450.1094P}).

\begin{figure*} 
\epsscale{1.}
\plotone{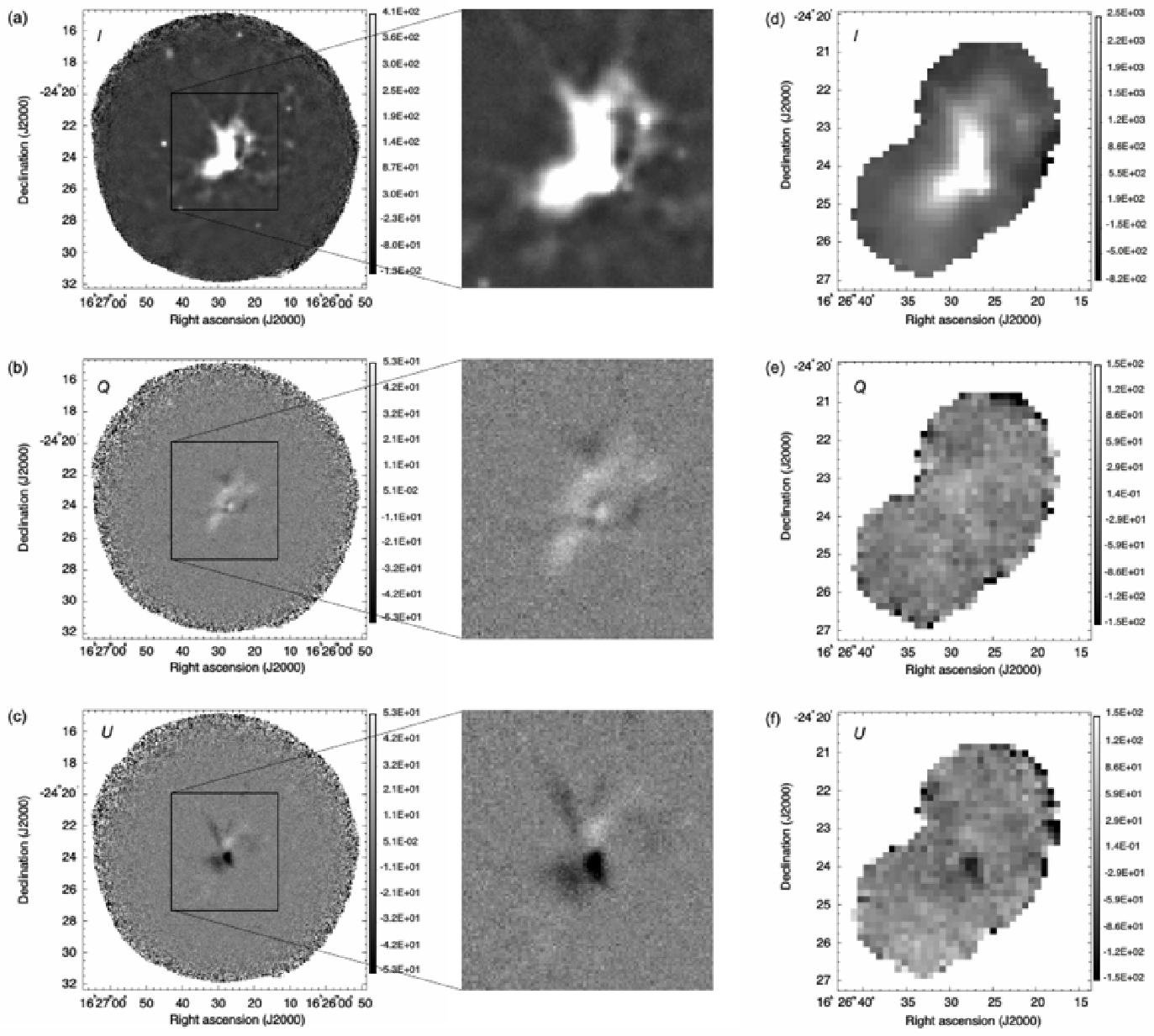}
\caption{
SCUBA-2/POL-2 data (this work) compared with SCUPOL (previous work).
The intensity grey scales are different between the POL-2 and SCUPOL data.
(a)--(c) Stokes $I$, $Q$, and $U$ images (linear scale) of the $\rho$ Oph-A core region obtained with the JCMT with SCUBA-2/POL-2.
The Stokes $Q$ and $U$ images (b and c) have the same grey scale in mJy beam$^{-1}$ units.
The SCUPOL field of view (d)--(f) is indicated by a black box
in Figures \ref{fig:StokesIQU}(a)--(c), respectively.
(d)--(f) 850 $\mu$m Stokes $I$, $Q$, and $U$ images of the $\rho$ Oph-A core region obtained from the JCMT with SCUPOL
(from The SCUBA Polarimeter Legacy Catalogue). 
The Stokes $Q$ and $U$ images (e and f) have the same grey scale
in mJy beam$^{-1}$ units.
The original units are pW and Volt in the SCUBA-2/POL-2 and SCUPOL data, 
which are converted to mJy using the conversion factors of 725 Jy beam$^{-1}$ pW$^{-1}$ and 455 Jy beam$^{-1}$ Volt$^{-1}$
(see \citealt{2009ApJS..182..143M}).
Note that the SCUPOL data are binned to generate 10$\arcsec$ vectors. 
The grey scale ranges in SCUBA-2/POL-2 and SCUPOL are different so that fainter regions in $Q$ and $U$ images are clearly seen in each image.
}
\label{fig:StokesIQU}
\end{figure*}

Figure \ref{fig:StokesIQU} shows a comparison between the SCUBA-2/POL-2 data and the previous polarization data 
from SCUPOL \citep{1999A&A...344..668G}. 
Figures \ref{fig:StokesIQU}(a)--(c) respectively show Stokes $I$, $Q$, and $U$ images of the $\rho$ Oph-A core region 
obtained from the JCMT with SCUBA-2/POL-2 (this work), and
Figures \ref{fig:StokesIQU}(d)--(f) respectively show Stokes $I$, $Q$, and $U$ images of the $\rho$ Oph-A core region 
obtained from JCMT with SCUPOL 
(previous work; see also the SCUBA Polarimeter Legacy Catalogue, \citealt{2009ApJS..182..143M}).
The black boxes in Figures \ref{fig:StokesIQU}(a)--(c) show the regions covered by SCUPOL.
As shown in Figure~\ref{fig:StokesIQU}, our data are deeper and more clearly provide the morphology of the surrounding regions
in all of the Stokes $I$, $Q$, and $U$ images, although two have the same spatial resolution.
However, it should be noted that the SCUPOL data are binned to generate 10$\arcsec$ polarization vectors\footnote{The polarization vectors are not true vectors since they give an orientation not a direction.}.

To compare the best intensity morphology with that from the previous data, 
we first introduce the detailed submillimeter morphology of the $\rho$ Oph-A core, and then present the polarimetric results.

\subsection{Morphology of $\rho$ Oph-A} \label{sec:morphology}

Figure \ref{fig:OphA_stokesI} shows the morphology of $\rho$ Oph-A from our 850 $\mu$m Stokes $I$ emission map.  
The region contains several sub-cores, which we outline below:  

\paragraph{Oph-A~SM1}
Oph-A SM1 is a sub-core located toward the peak of the 850 $\mu$m intensity (\citealt{1989MNRAS.241..119W}, also cf. Figure \ref{fig:OphA_stokesI}).  
It has the brightest submillimeter continuum in all of $\rho$ Oph. 
The filamentary morphology in $\rho$ Oph suggests that SM1 may be influenced by the B4 star Oph~S1 (cf. Figure \ref{fig:OphA_stokesI}), 
which is a nearby young B-type star. 
\citet{1998A&A...336..150M} report that the total mass and dust temperature of Oph-A~SM1 are 2 $M_{\sun}$ and $T \approx 20$~K, respectively.

\paragraph{VLA~1623}
VLA~1623 is the prototypical Class 0 star \citep{1993ApJ...406..122A}.  
It drives a large-scale bipolar molecular outflow (\citealp{1995MNRAS.277..193D, 1997ApJ...479L..63Y}) and is embedded within a nearly spherical dust envelope \citep{1993ApJ...406..122A}.
\citet{1997IAUS..182P..63B} found 
three emission clumps at centimeter wavelengths with the Very Large Array, which they interpreted as knots in the radio jet driving the large CO outflow (see also, \citealt{2013ApJ...768..110C}).
However, the position angles of the radio jet and the CO outflow differ by approximately 30\arcdeg.
Clump~A was further resolved into two components at a high angular resolution \citep{2013ApJ...768..110C},
with the Submillimeter Array (SMA). 
VLA~1623 is also binary system, with two components separated at high angular resolutions \citep{2000ApJ...529..477L, 2011MNRAS.415.2812W}.  
Since the POL-2 resolution is approximately 14\farcs1 at 850 $\mu$m, 
we cannot separate these components and refer to them as a single source, VLA 1623 in this paper.

\paragraph{Other local structures}
There are two filaments in the north part of the $\rho$ Oph-A core. 
These structures are consistent with not only the results obtained with SCUBA on the JCMT  \citep{1999ApJ...513L.139W}
but also those seen in the map made with SCUBA-2 \citep{2015MNRAS.450.1094P} and IRAM \citep{1998A&A...336..150M} results.
In addition to the filaments, \cite{1999ApJ...513L.139W} reported that 
there are two arcs of emission in the direction of the northwest extension of the VLA 1623 outflow. 
The outer arc appears relatively smooth at 850 $\mu$m,
while the inner arc breaks up into a number of individual clumps,
some of which are known protostars.

\subsection{New Submillimeter Polarization Vector Map} \label{sec:vectormap}

Polarized thermal emission from dust grains in clouds 
offers an ideal probe of the magnetic field structure 
on multiple scales, from protostellar disks to cores and clumps
(e.g., \citealp{2001ApJ...562..400M, 2004ApJ...600..279C}).

\begin{figure*} 
\epsscale{.7}
\plotone{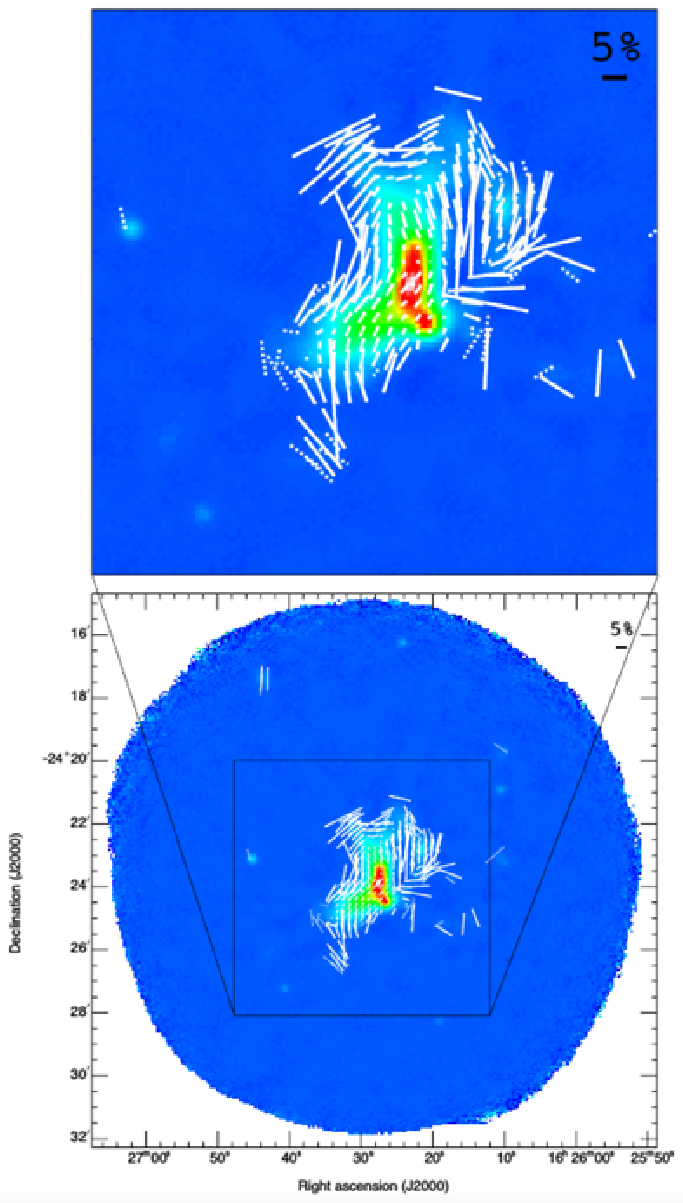}
\caption{
The 850 $\mu$m polarization vector maps. 
The vectors are sampled on a 12$\arcsec$ grid ($3 \times 3$ pixel smoothing)
and plotted where $I>0$, $P/\delta P > 2$, and $\delta P < 4\%$ (dotted vectors)
and $I>0$, $P/\delta P > 3$, and $\delta P < 4\%$ (solid vectors), respectively. 
}
\label{fig:submm}
\end{figure*}

Figure \ref{fig:submm} shows our submillimeter polarization vector maps of the $\rho$ Oph-A region, 
observed with SCUBA-2/POL-2. 
Since it is worthwhile to directly compare our data with the previous submillimeter polarization vector map,
Figure \ref{fig:submm} is prepared with the same criteria, $I>0$, $P / \delta P > 2$, and $\delta P < 4\%$,
as were used in the previous results by SCUPOL (See Figure 44 of \citealt{2009ApJS..182..143M}).
The selection criteria here are mainly for the purpose
of the comparison with the SCUPOL data; however,
we have found this to be fairly reasonable to see the magnetic field 
structure in this region by changing various $P/\delta P$ or $\delta P$ selections.
In addition, Figure~\ref{fig:submm} suggests the vector maps with both $P / \delta P > 2$ and $P / \delta P > 3$ 
are almost the same, if we use the additional criterion of $\delta P < 4\%$.
Without this $\delta P$ criterion, the vector map with $P / \delta P > 2$ 
has many more vectors, but has RMS noise values of $\delta P$ and $\delta \theta$ too high to allow interpretation of the magnetic field behavior.
Therefore, we use the criteria of  $I > 0$, $P / \delta P > 2$, and $\delta P < 4\%$ in the following discussion
to maximize the number of polarization vectors that can be used for our discussion below on the magnetic field directions. 
The angle errors ($\delta\theta$) of approximately 15$\arcdeg$ in the 2$\sigma$ case are acceptable for such discussions. 
Therefore, we show both 2$\sigma$ and 3$\sigma$ cases in several figures and use 2$\sigma$ data for the discussion on magnetic field directions.

Our data are more sensitive than those
obtained by SCUPOL as shown in Section \ref{sec:verification}.
The new submillimeter polarization vectors inside the dense regions agree well with the results by \citet{2009ApJS..182..143M},
especially in the bright region near SM1.
The dominant submillimeter polarization position angle in the bright region is approximately 130$\arcdeg$ (as discussed in the following section).
We have also checked if our new data are consistent with the JCMT 800 $\mu$m aperture polarimetric data of \citet{1996A&A...309..267H}. 
The measured positions are not exactly the same, but both $P$ and $\theta$ values are consistent with each other
between the two studies.

Note that there are clear 
inconsistencies between SCUBA-2/POL-2 and SCUPOL data 
in the outer parts of the Oph-A core region.
To the south-east and south of the core, SCUPOL data show more numerous polarization vectors 
even at a very low intensity level 
while to the north-west  and north-east, SCUBA-2/POL-2 data reveal more vectors.
Since our data have a higher S/N ratio, as shown in Figure \ref{fig:StokesIQU}, 
we believe that our SCUBA-2/POL-2 data are more reliable in the faint outer regions,
down to approximately 30 mJy beam$^{-1}$,
while care is necessary when using the SCUPOL vectors in the outer regions.
The reason why the SCUPOL data have more vectors in some outer core regions is not clear. However, we note that the SCUPOL maps were made by chopping, while POL-2 in a scanning mode. Therefore, the chopping effect cannot be excluded in the SCUPOL data, which were taken at different times by different observers.
\begin{figure} 
\epsscale{1.}
\plotone{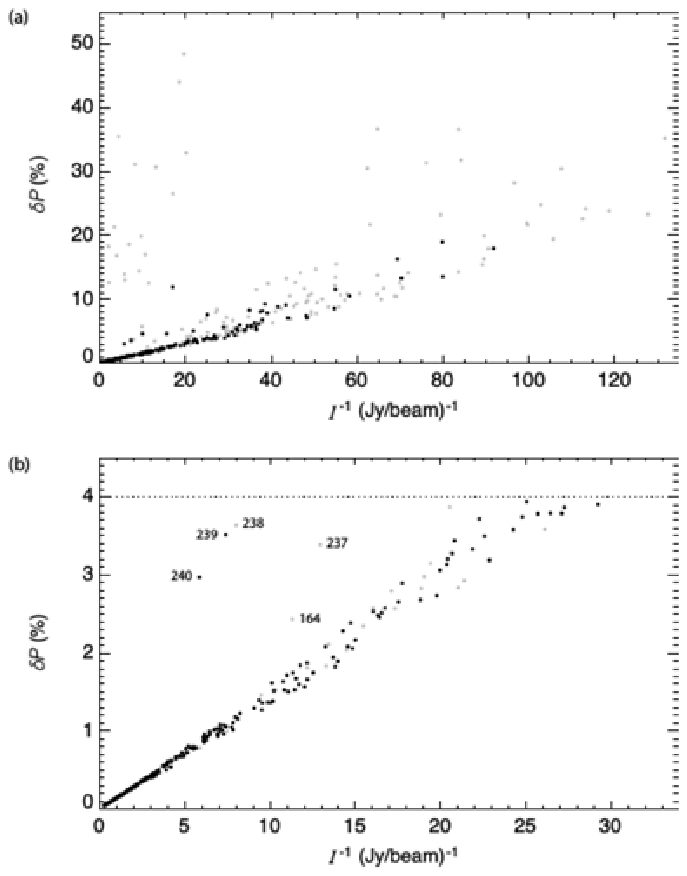}
\caption{
Degree of polarization errors ($\delta P$) vs. 1/Intensity ($I$).
Five dots with large $\delta P$ are labeled (cf. Table 1). 
These five sources are located in outer regions of the Oph-A core region where the noise levels are higher. 
(a) Grey circles: $I>0$, $P/\delta P > 2$. Black circles: $I>0$, $P/\delta P > 3$. 
(b) Grey circles: $I>0$, $P/\delta P > 2$, and $\delta P < 4\%$. Black circles: $I>0$, $P/\delta P > 3$, and $\delta P < 4\%$.
}
\label{fig:plot}
\end{figure}

Based on the robustness toward the fainter regions mentioned above, our data clearly show the polarizations in the fainter regions surrounding the core and {the degrees of polarization are much higher ($> 5\%$) in the outer envelope. 
This trend is clear in our polarization map whose vector length is proportional to the degree of polarization
(see also Figure~\ref{fig:plot2}).

\subsection{Experimental Criteria}

Figure \ref{fig:plot}  shows
the degree of polarization errors ($\delta P$) versus the inverse intensity ($I^{-1}$);
the polarization uncertainty increases steadily with decreasing intensity.  
For $\delta P > 4\%$, we see significant scatter in this relation, whereas the data with $\delta P < 4\%$ are fairly well correlated.  
There are five vectors with $\delta P < 4\%$ that show substantial scatter from the main trend and are labeled on Figure 4 (cf. Table 1).  
Aside from these five cases, vectors with $\delta P < 4\%$ appear to be robust.   
All five anomalous positions (ID: 164, 237, 238, 239, 240) are located 
near the map boundary where the noise levels are higher.
Vector 164 is located in the east of Oph-A, and
vector 237 is located between the north-west filamentary structure and GSS 26 and
vectors, 238, 239 and 240, are located in the upper part of the north-east filamentary structure.
Note that including these five vectors does not affect 
our results.
Figure~\ref{fig:plot} also shows that our polarization data 
present a large scatter when the intensity levels are less than
approximately 30~mJy beam$^{-1}$, 
which corresponds to $N$(H$_2$) $\sim$ 4 $\times$ 10$^{21}$ cm$^{-2}$  assuming a temperature of 10~K \citep{2007PhDT.......435K}.

\section{DISCUSSION} \label{sec:discussion}

Magnetic fields in star formation are significant as they can influence core collapse, star formation rates, and molecular cloud lifetimes 
(e.g., \citealp{1988ApJ...329..392M, 2001ApJ...562..852H, 2000ApJ...539..342E}).   
We use our polarization data to determine the magnetic field strength in $\rho$ Oph-A below.

\subsection{Magnetic Field Structures in $\rho$ Oph-A} \label{sec:mf}

The $\rho$ Oph molecular cloud has been observed with 1.3~mm continuum mapping \citep{1998A&A...336..150M} 
and line mapping (\citealt{1999ApJ...525L.105U}, see also \citealt{2015MNRAS.447.1996W}), 
and the $\rho$ Oph-A core region is one of the most obvious sources.
\citet{2009ApJS..182..143M} presented a bulk analysis of SCUPOL 850~$\mu$m polarization vector maps, 
which include the $\rho$ Oph-A core.
The submillimeter polarization position angle is about 130$\arcdeg$ in average
(measured east of north),
which indicates a magnetic field direction of approximately 40$\arcdeg$ (by rotating the submillimeter polarization vectors by 90$\arcdeg$).
This angle is consistent with the well-known 50$\arcdeg$ component determined via infrared polarimetry observations
\citep{1988MNRAS.230..321S, 2015ApJS..220...17K}.
Therefore, the magnetic field seems largely consistent between the outer low-density cloud and the high density cores.

\begin{figure*} 
\epsscale{1.}
\plotone{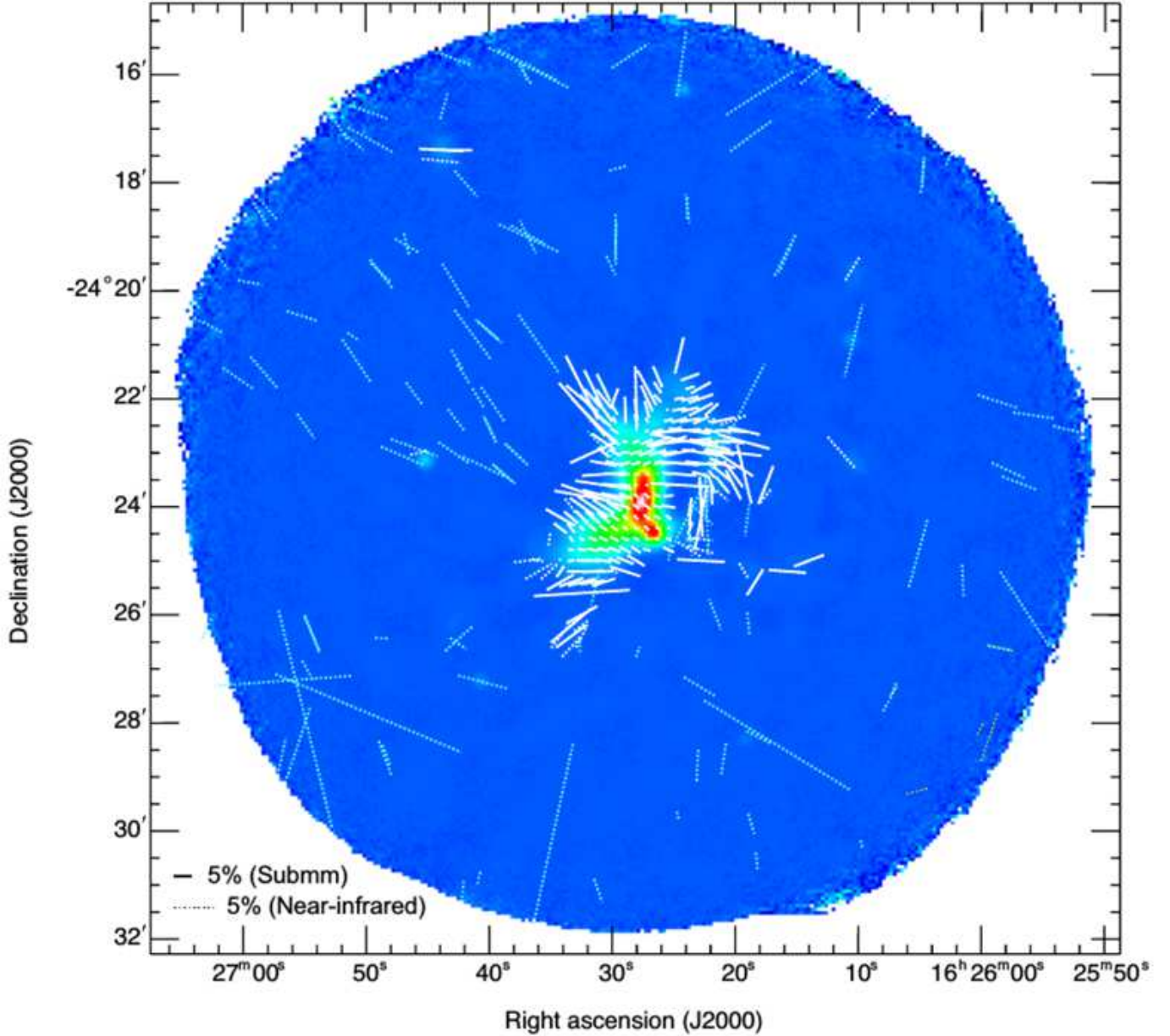}
\caption{
Same as Figure \ref{fig:submm} but rotated by 90$\arcdeg$ with
near-infrared polarization vectors (cyan dotted vectors) from \citet{2015ApJS..220...17K}.
These vectors therefore show the inferred magnetic field orientation projected on the plane of the sky.
Scale vectors of 5\% at submillimeter and near-infrared wavelengths are shown in the bottom left corner.
}
\label{fig:submm90}
\end{figure*}

To investigate magnetic field structures in this region in more detail,
we use the POL-2 polarization vectors rotated by 90$\arcdeg$, as shown in Figure \ref{fig:submm90}.
Figure \ref{fig:submm_mf} shows the inferred morphology of the magnetic field 
in the $\rho$ Oph-A core region.
In this figure, vector maps are shown in two ways; one selected with the
polarization signal-to-noise ratio ($P/\delta P>$ 2 or 3) and the other selected with
the intensity signal-to-noise ratio ($I/\delta I>20$). 
The latter intensity-selection is shown  
because selecting by S/N in $P$ will tend to bias the polarization data sample to high $P$ values
(especially towards regions of low intensity), 
so a comparison with a sample selected by $I/\delta I$ is made to show that without this bias
the polarization fraction is still larger on average for cloud sightlines in the envelope.
Figure \ref{fig:plot2} demonstrates that this correlation is robust,
as seen also from the negative correlation between the degrees of polarization
and intensities in both of the $P$ and $I$ selection data.
We find that $P \propto I^{\gamma}$ where $\gamma \sim -0.8$ for the $P$ selection to $-0.7$ for the $I$ selection.
Note that there is a larger dispersion at the low-intensity regions in the $I$ selection 
because not only high $P$ data but also several low $P$ data exist in the $I$ selection.
This trend might be due to a combination of several factors such grain alignment and magnetic field geometry. 
A detailed discussion will be presented elsewhere.
\begin{figure} 
\epsscale{1.}
\plotone{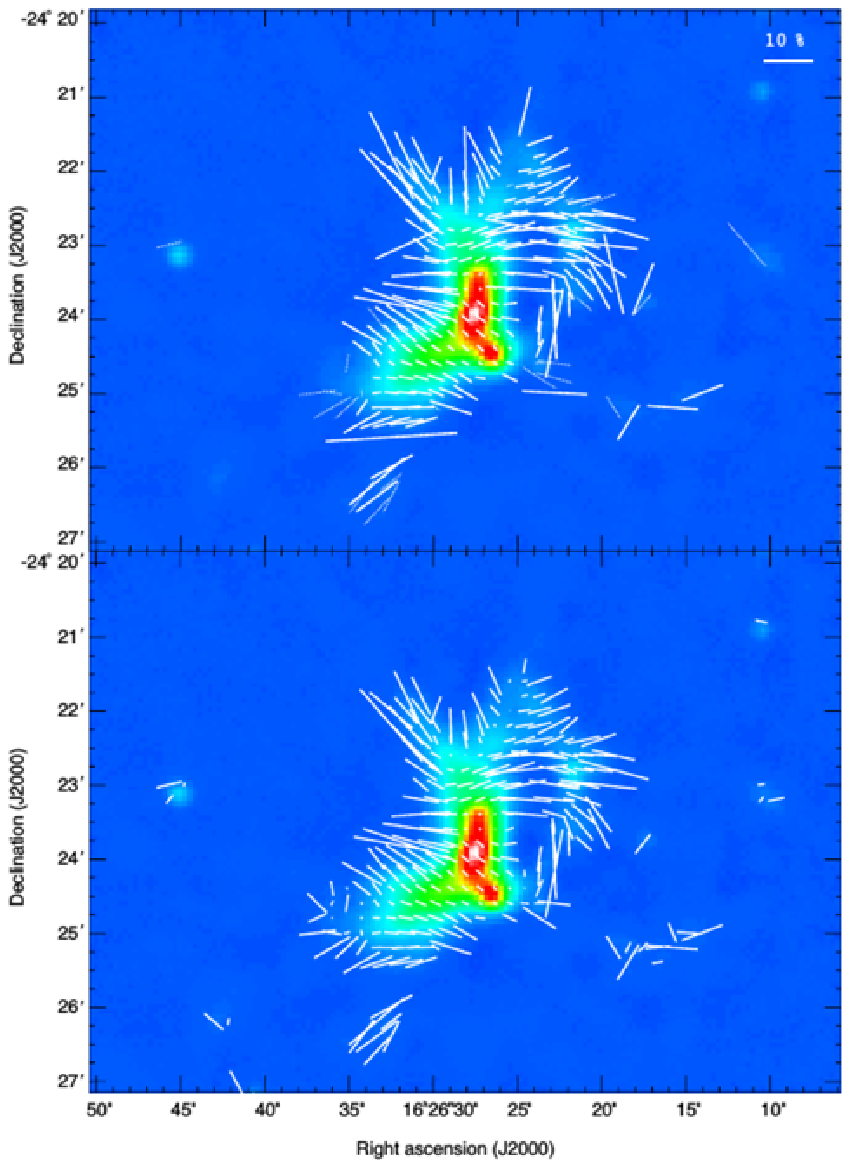}
\caption{
The 850 $\mu$m polarization vector maps, sampled on a 12$\arcsec$ grid and rotated by 90$\arcdeg$ 
(i.e., as in Figure \ref{fig:submm} but rotated by 90$\arcdeg$).
The 90$\arcdeg$ rotated vectors, which 
show the inferred magnetic field orientation projected on the plane of the sky,
are plotted where 
$I>0$, $P/\delta P > 2$, and $\delta P < 4\%$ (dotted vectors) and
$I>0$, $P/\delta P > 3$, and $\delta P < 4\%$ (solid vectors)
in the top panel
and where $I>0$ and $I/\delta I > 20$ in the bottom panel, respectively.
A 10\% scale vector is shown in the upper right corner.
See text for these two panels with different selections ($P/\delta P$ and $I/\delta I$).
}
\label{fig:submm_mf}
\end{figure}
\begin{figure} 
\epsscale{1.}
\plotone{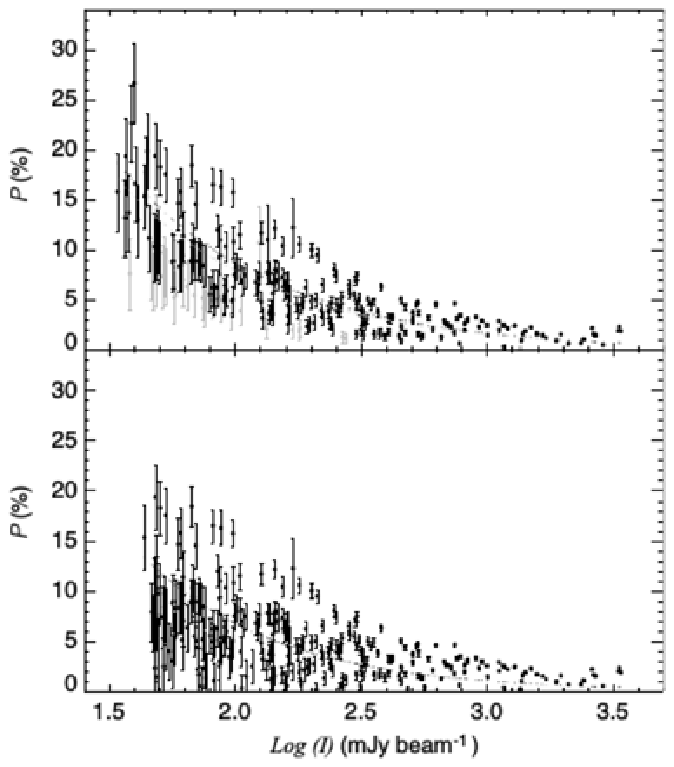}
\caption{
Degree of polarization ($P$) vs. Intensity ($I$).
Top panel: Black circles show the sources with $I>0$, $P/\delta P > 3$, and $\delta P < 4\%$, and 
                  grey circles show the sources with $I>0$, $P/\delta P > 2$, and $\delta P < 4\%$.
Bottom panel: Black circles show the sources with $I>0$, $I/\delta I > 20$, and $P > 0$.
Each least-square-fit power-law is shown as dotted curves.
}
\label{fig:plot2}
\end{figure}

We have found by eye that there are at least 10 distinct magnetic field components 
in the core region, and we refer to them as ``Components a--j'' 
(see Figure \ref{fig:region}).
Please note that our division of these components does not
mean that these field component are always independent
but all or some of them could be smoothly connected with eath other.
The aim of the region division here is mainly to identify 
the change of directions and degrees of the polarization vectors
and to compare them with the near-infrared polarization data.
A summary of these components is as follows:

\begin{enumerate}[label=(\alph*)]
 \setlength{\itemsep}{-5pt}
\item  
Small $P$ ($<$ 3\%) and $\sim$50$\arcdeg$ component at SM~1 and VLA~1623 
	around the center of the observed field-of-view,
\item  
Large $P$ and $\sim$40$\arcdeg$ component near A-MM~7 to the east of A-MM~5, 
\item  
Large $P$ and $\sim$20$\arcdeg$ component near A-MM~5, 
\item  
Large $P$  and $\sim$100$\arcdeg$ component at A-MM~4, 
\item  
Small $P$ ($<$ 3\%)   and $\sim$100$\arcdeg$ component between A-MM~5 and SM~1N, 
\item 
Large $P$ and $\sim$80$\arcdeg$ component to the west of SM~1, 
\item  
Large $P$ and $\sim$70$\arcdeg$ component to the east of SM~1 and SM~1N, 
\item  
Large $P$ and $\sim$80$\arcdeg$ component at A-MM~3, 
\item 
Small $P$ ($<$ 3\%)  and $\sim$80$\arcdeg$ component between SM~2 and A-MM~8, 
\item 
Large $P$ and $\sim$120$\arcdeg$ component between SM~2 and A-MM~8. 
\end{enumerate}

Figure \ref{fig:region} illustrates that these components differ from each other either in polarization position angle or degree of polarization 
(see also Table \ref{tab:region}).
Components a, c, and i are already seen in, and are consistent with, SCUPOL data \citep{1999sf99.proc..212T}. 
Components b, d, e, f, g, h, and j are additionally identified in our SCUBA-2/POL-2 data.
One can also see the polarization vectors associated with the Components b, d, e, g, and j in \citet{2009ApJS..182..143M}.
We also note that our results suggest the magnetic field 
is mostly well organized 
(rather than disordered due to turbulence; see Section \ref{sec:MFstrength}).
\begin{figure*} 
\epsscale{.9}
\plotone{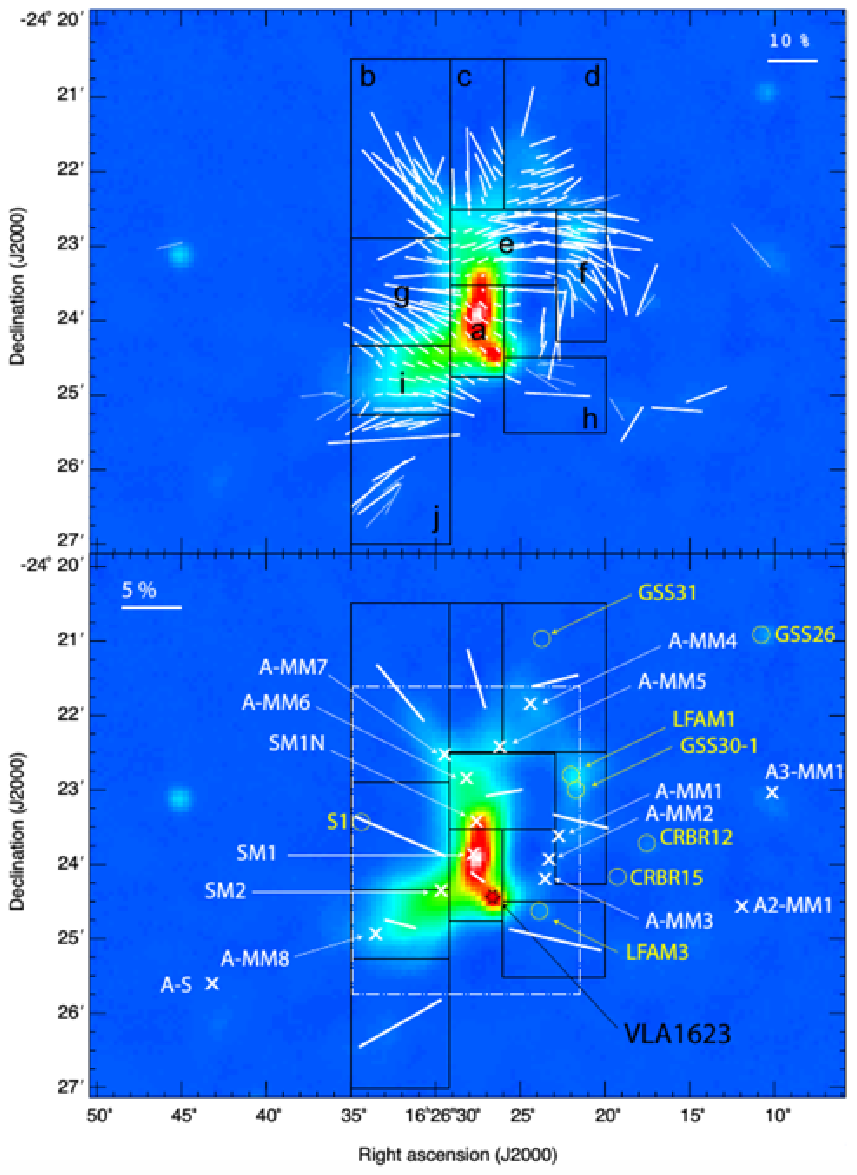}
\caption{
Close up of the $\rho$ Oph-A core region in Figure \ref{fig:OphA_stokesI},
with regions divided according to the magnetic field direction (top panel), and
showing the median of the magnetic field directions of each region shown in Table 2 (bottom panel).
Black boxes indicate the Componets a-j (see Section \ref{sec:mf}).
10\% and 5\% scale vectors are shown in the upper right (top panel) and left (bottom panel) corners, respectively.
Note that $P/\delta P > 2$ (dotted vectors) and $P/\delta P > 3$ (solid vectors) data are shown here, 
therefore the errors of the polarization vector angle are typically much less than 15$\arcdeg$.
The region indicated with the white dash-dotted box in the bottom panel is corresponding to the region
shown in Figure \ref{fig:velocity}.
}
\label{fig:region}
\end{figure*}

In the central region around SM~1 (Component a),
the vectors are well aligned with the 50$\arcdeg$ 
magnetic field component observed in the surrounding medium on various scales
(see Section \ref{sec:tracing}).
Although the average direction of the main component is approximately 50$\arcdeg$, 
the magnetic field tends to be locally perpendicular (approximately 100$\arcdeg$--110$\arcdeg$) to the arc-structure
(south part of Region f).
Between SM~1N and A-MM~6, the magnetic field direction is almost east-west (Component e)
while the arc extends to the north-east or north-west, and the magnetic field directions extend toward LFAM1 and GSS 30--1 
(Component f).
A perpendicular field relative to the core shape (i.e., the elongation of the arc-structure between north-east and north-west filaments) 
is important 
for the formation and growth of this core.
Such orthogonal fields are often seen in the densest parts of the cloud or cloud cores
(\citealp{1987MNRAS.224..413T, 1988MNRAS.231..445T, 1998ApJ...506..306N, 2013A&A...550A..38P, 2014ApJ...784..116M, 2016ApJ...824..134F, 2013ASPC..476...95A, 2016A&A...586A.138P}).

There are other local structures besides the 50$\arcdeg$ component.
To the north of A-MM6 (Component c), the magnetic field direction is almost north-south, and
to the north of A-MM7, the magnetic field direction is almost north-east (Component b), 
which is the same as the direction of the north-east filament.
Notable are the low degree of polarization near SM~1N 
and some deviation in magnetic field direction near VLA~1623 and its outflow region.

In this paper, we have assumed that the 850 $\mu$m emission measured by SCUBA-2
is dominated by the thermal dust continuum emission. 
However, 
the continuum emission can be contaminated by CO (3--2) line emission \citep{2012MNRAS.426...23D}.
Figure \ref{fig:CO} shows an overlay of the CO (3--2) line emission from \citet{2015MNRAS.447.1996W} on the 850 $\mu$m continuum map. 
In the dense center of Oph-A, the CO contamination fraction is typically $< 1 \%$. 
However, in the brightest regions of CO emission from the outfow from VLA~1623,
the contamination fraction can be much higher \citep{2015MNRAS.450.1094P}.
The regions that have high CO contamination have very low column density values, 
and are mostly along the jet axis between Oph-B and Oph-C/E/F, outside our field of view.  
In the dense center of Oph-A, 
the fractional contribution of CO is $< 1 \%$ and generally does not exceed 10\% 
anywhere on source.
The ring-shaped region seen in our Stokes $I$ image to the west of Oph-A 
is dominated by the CO emission rather than the thermal dust emission.
Since the CO emission is weak toward the bright dust emission ($< 5\%$), 
even if it is polarized by the Goldreich-Kylafis effect, it will contribute minimally to our results.
\begin{figure*} 
\epsscale{1.}
\plotone{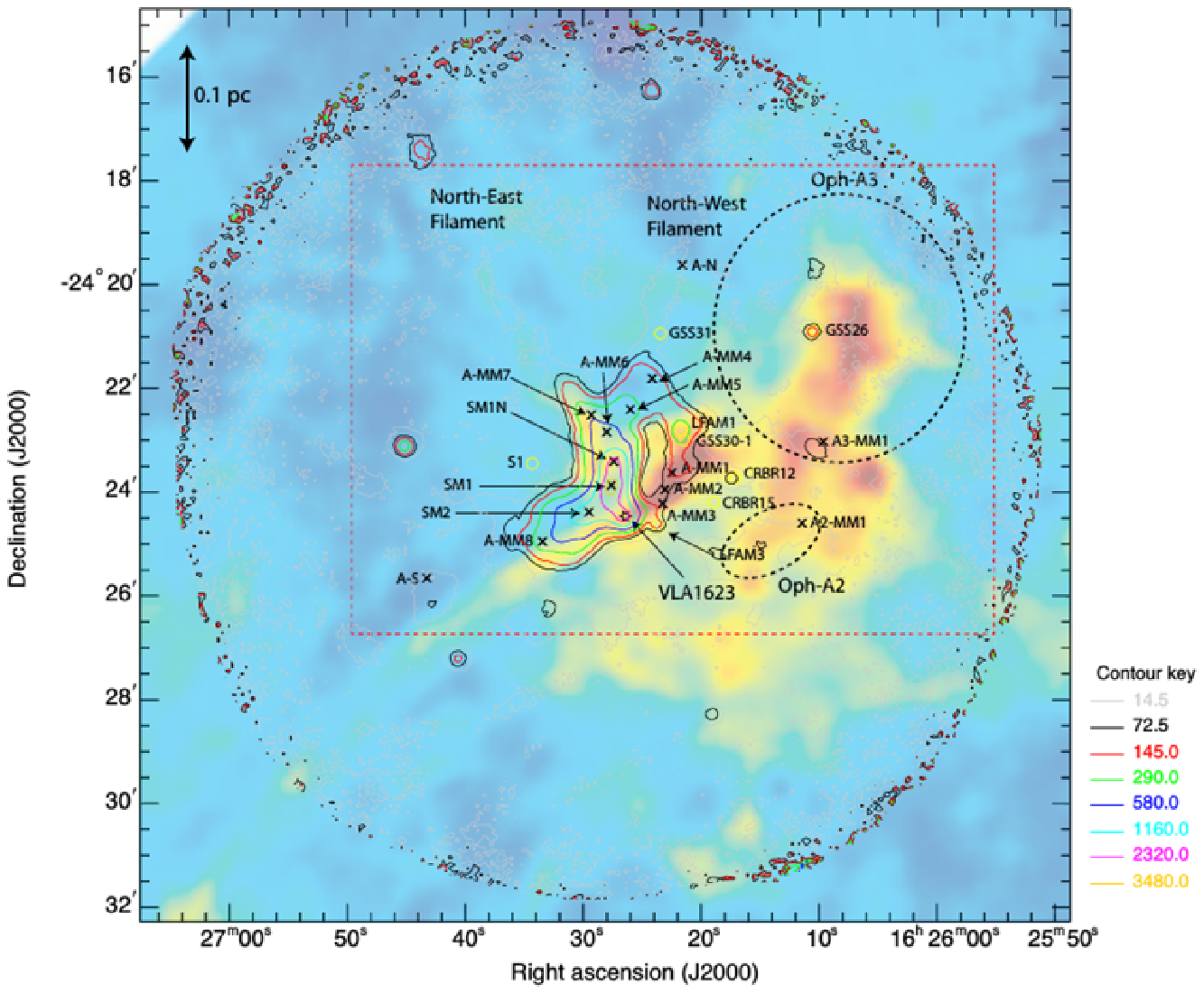}
\caption{
Contours of the 850 $\mu$m total intensity (Stokes $I$) image (cf. Figure \ref{fig:OphA_stokesI} of this work) of the $\rho$ Oph-A field 
superimposed on HARP CO $J = 3-2$ observations showing the integrated emission between $-$5 and $+$12 km s$^{-1}$ 
(cf. Figure 1 of \citealt{2015MNRAS.447.1996W}).
}
\label{fig:CO}
\end{figure*}

\subsection{Local Magnetic Field Strength} \label{sec:MFstrength}

Polarization arising from dust grains, which are aligned with their major axes perpendicular to the magnetic field (e.g., \citealt{2009ApJ...697.1316H}),
allows us to estimate the magnetic field direction.
However, the present uncertainties in theories of dust grain alignment 
limit the ability with current techniques to trace magnetic fields without ambiguities 
(see \citealp{2007JQSRT.106..225L, 2015psps.book...81L}, for a review). 

The most common method to infer a magnetic field strength from polarized dust emission is the 
Davis-Chandrasekhar-Fermi method (more commonly referred as the Chandrasekhar-Fermi (CF) method;
\citealt{1951PhRv...81..890D, 1953ApJ...118..113C}; see also \citealt{2016ApJ...820...38H} and \citealt{2017ApJ...846..122P}).
The Davis-Chandrasekhar-Fermi method infers a magnetic field strength by
statistically comparing the dispersion in the polarization orientation with the dispersion in velocity.
Therefore, the magnetic field strength projected on the plane of the sky can be calculated by
\begin{equation}
\begin{split}
   B_p &= {\cal Q} \sqrt{4\pi\rho}\ {{\delta v_{\rm los}}\over{\delta\theta}} \\
\end{split}
   \label{eq:CFmethod}
\end{equation}
assuming that velocity perturbations are isotropic \citep{2001ApJ...546..980O}.
In Equation (\ref{eq:CFmethod}), $\cal Q$ is a factor to account for various averaging effects 
(see \citealt{2004ApJ...616L.111H} and \citealt{2004ApJ...600..279C} for details),
$\rho$ is the mean density of the cloud, 
$\delta v_{\rm los}$ is the rms line-of-sight velocity, 
and $\delta\theta$ is the dispersion in the polarization angle. 
To estimate a magnetic field strength in the $\rho$ Oph-A core region,
a correction factor of $\cal Q=$0.5 
is adopted here because the magnetic field appears to be ordered (\citealt{2001ApJ...546..980O, 2004ApJ...616L.111H}, also see \citealt{2008ApJ...679..537F, 2009ApJ...695.1362N, 2017ApJ...846..122P}).
Since we apply this formula only to the sub-regions where the angle dispersion is relatively small ($\leq 25\arcdeg$)
and the velocity dispersion of the molecular lines tracing high density regions is available in the next paragraph,
$\cal Q$ = 0.5 is appropriate, as simulated by \citet{2001ApJ...546..980O}.
Note that if the turbulence correlation length is not resolved and therefore their simulation assumption is not valid,
the case the $\cal Q$ factor can be much lower (\citealt{2001ApJ...561..800H}, see also \citealt{2009ApJ...706.1504H}).
Then Equation (\ref{eq:CFmethod}) can be expressed as follows \citep{2002ApJ...566..925L}: 
\begin{equation}
\begin{split}
      B_p &= 8.5 {\sqrt{n_{\rm H2}/(10^6 \, \rm cm^{-3})} \, \Delta v / (\rm km \, s^{-1}) \over{\delta\theta (\arcdeg)}} \rm \, mG
\end{split}
   \label{eq:CFmethod2}
\end{equation}
Here, $n_{\rm H_2}$ is the number density of hydrogen molecules  
and $\Delta v$ is the line width.

As mentioned in previous section, there are several magnetic field components in the $\rho$ Oph-A core region.
Since they are different from each other either in direction or degree of polarization,
we estimate the magnetic field strength of each 
component separately. 
To investigate their magnetic field strengths individually, 
we estimated median polarization position angles, 
which indicate the local average magnetic field directions of each component. 
Table \ref{tab:region} shows the median degrees of polarizations and position angles
calculated using Stokes $Q$ and $U$ in each region. 
Figure \ref{fig:region} shows the vectors in each region averaged over in each region. 

The local average density in each of the Components a--j
is calculated from our Stokes $I$ data, 
assuming that the core-depth is equal to the geometric mean size of each sub-core where the polarization data exist,
and ranges from $2 \times 10^6$ cm$^{-3}$ to $7 \times 10^4$ cm$^{-3}$.
Since the $\rho$ Oph-A core has a complex magnetic field structure showing various directions
in each sub-core, we do not attempt to apply to
the Davis-Chandrasekhar-Fermi method to the entire core 
but only to the sub-cores showing a relatively well-defined
magnetic field direction.
In Components a, d and e, 
\cite{2007A&A...472..519A} estimated
a velocity dispersion of 0.26 km~s$^{-1}$, 0.15 km~s$^{-1}$, and 0.17 km~s$^{-1}$, respectively.
These are non-thermal line dispersions from N$_2$H$^+$ (1--0) observations.
Using these values with the standard deviation in field direction of 1.5$\arcdeg$, 5.8$\arcdeg$, and 2.7$\arcdeg$
found in each region, the magnetic field strength projected on the plane of the sky is calculated as $Bp \sim$ 
5, 0.2, and 0.8 mG (cf. Table \ref{tab:region}). 
The estimated magnetic
field strength in the $\rho$ Oph-A core region is larger 
than that in other molecular clouds derived using the Davis-Chandrasekhar-Fermi method (namely 20--200 $\mu$G; \citealt{2005MNRAS.356.1088A, 2006ApJ...650..945P, 2008A&A...486L..13A, 2010ApJ...708..758K, 2011ApJ...741...35K,
2011ApJ...734...63S, 2015ApJ...798...60K}) but comparable to those in the
Orion~A region (see e.g., \citealt{2017ApJ...846..122P}). 
Our high magnetic field strengths may be attributed to using the higher H$_2$ densities 
associated with the sub-cores rather than the lower H$_2$ densities associated with the larger $\rho$ Oph-A core.  
Thus, we conservatively take these field strengths as an order-of-magnitude estimate.  
These values are still representative of the field strength toward the sub-cores in $\rho$ Oph-A 
and can be taken as an upper limit for the surrounding gas.

Finally, it should be noted that there are certain limitations in the Davis-Chandrasekhar-Fermi technique
such as the effect of the limited telescope resolution \citep{2001ApJ...561..800H}.
Also note that our estimates are only for some sub-regions
where the field dispersions are relatively small. Therefore, 
both of these effects tend to bias towards a high magnetic field strength.
In addition, more sophisticated applications of 
the Davis-Chandrasekhar-Fermi technique such as 
described in \citet{2017ApJ...846..122P} or \citet{2009ApJ...696..567H}
would be desirable in future works.

\subsection{Magnetic Fields and Centroid Velocity} \label{sec:velocity}

Our polarimetric data will be useful to discuss the correlation
between the magnetic field and the velocity field in each core. 
However, this will be beyond the scope of this first-look paper.
Therefore, in this section, we show an example of a possible correlation
between magnetic field and velocity gradient.

Strong Alfv\'enic turbulence develops eddy-like motions perpendicular to the local magnetic field direction \citep{1995ApJ...438..763G}. 
Very recently, \cite{2017ApJ...835...41G} 
have proposed that 
this fact can be used to study the direction of magnetic fields by using the velocity gradient 
calculated from the centroid velocity.
The centroid velocity is an intensity-weighted average velocity along the line of sight (e.g., \citealt{1999ApJ...524..895M}).
Here, we try to compare the magnetic field direction in the $\rho$ Oph-A core region with centroid velocity components.

\cite{2007A&A...472..519A} measured subsonic or transonic levels of internal turbulence 
within the condensations, 
and their result supports the view that
most of the L1688 starless condensations are gravitationally bound and prestellar in nature.
Figure \ref{fig:velocity} shows a comparison between magnetic field direction  (this work) and centroid velocity components
of N$_2$H$^+$(1--0) spectra \citep{2007A&A...472..519A}.
The apparent main velocity core gradient (indicated by arrows in Figure \ref{fig:velocity}) appears to be roughly perpendicular to the magnetic field orientation traced by POL-2. Certainly these observations should be compared with theoretical modeling using the physical parameters of the $\rho$ Oph-A core in future.

\begin{figure} 
\epsscale{1.}
\plotone{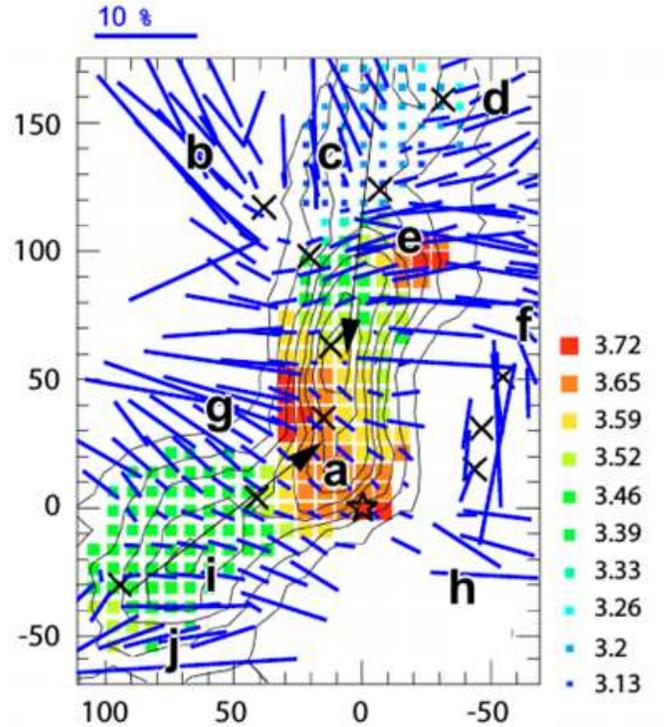}
\caption{
Comparison between magnetic field directions from this work
and the centroid velocity components
of N$_2$H$^+$(1--0) spectra (filled squares of varying sizes and colors,
Figure 6(c) of \citealt{2007A&A...472..519A}).
The (0, 0) offset corresponds to the position 
$\alpha$ = 16$^{\rm h}$ 26$^{\rm m}$ 26\fs45, $\delta$ = -24$\arcdeg$ 24$\arcmin$ 30.8$\arcsec$ [J2000].
The contours, which were drawn by \citet{2007A&A...472..519A}, go from 2 to 16 K km s$^{-1}$. 
The color code shows the velocity centroid.
The underlying contours represent the same N$_2$H$^+$(1--0) integrated intensity maps.
Our suggested centroid velocity gradient is shown by black arrows,
which are not in the original figure \citep{2007A&A...472..519A}.
Shown in the upper left is the 10\% scale vector for the 90$\arcdeg$-rotated submillimeter polarization vectors.
The letters a--j are the magnetic field components defined from our submillimeter polarimetry (see text).
Crosses mark the 1.2 mm continuum positions of starless condensations, while stars mark the positions of VLA~1623 
(see \citealt{2007A&A...472..519A}).
The labels a--j indicate the distinct magnetic field components in each sub-region a--j.
}
\label{fig:velocity}
\end{figure}

\subsection{Tracing magnetic fields across different wavelengths} \label{sec:tracing}

Polarimetry in the $\rho$ Oph-A core region was reported previously by several authors at other wavelengths.
\citet{1988MNRAS.230..321S} carried out near-infrared polarimetry (in the $K$ band only) of 20 sources
which are embedded within the densest region of the $\rho$ Oph dark cloud 
with a single channel detector, 
and they suggested that there are three dominant components of the polarization position angles 
0\arcdeg, 50\arcdeg, and 150\arcdeg. 
Recently, \citet{2015ApJS..220...17K} presented wide and deep near-infrared polarimetry (in the $JHK_s$ bands) 
of the $\rho$ Oph regions,
which corresponds to the densest part of L1688. 
Since they cover a wider region than our observations
(but much more sparsely due to the limited number of stars
available for the aperture polarimetry),
we compare their polarimetry data covering a 40\arcmin $\times$ 40\arcmin region with our submillimeter data.
In this active cluster-forming region,
they found that the magnetic fields appear to be connected from core to core, 
rather than as a simple overlap of the different cloud core components. 
Putting it differently, the magnetic field morphology seems to be connected between different cores 
in the $\rho$ Oph molecular cloud complex. 
In addition, 
comparing their near-infrared polarimetric results 
with the large-scale magnetic field structures obtained from previous optical polarimetric study \citep{1976AJ.....81..958V}, 
they suggested that the magnetic field structures in the $\rho$ Oph core were distorted by the cluster formation in this region, 
which may have been induced by shock compression due to wind/radiation from the Scorpius--Centaurus association.
Also note that there is 350 $\mu$m submillimeter polarization from the CSO in \citet{2010ApJS..186..406D}  for $\rho$ Oph-A.
Their data are broadly consistent with our 850 $\mu$m map.

\begin{figure} 
\epsscale{1.}
\plotone{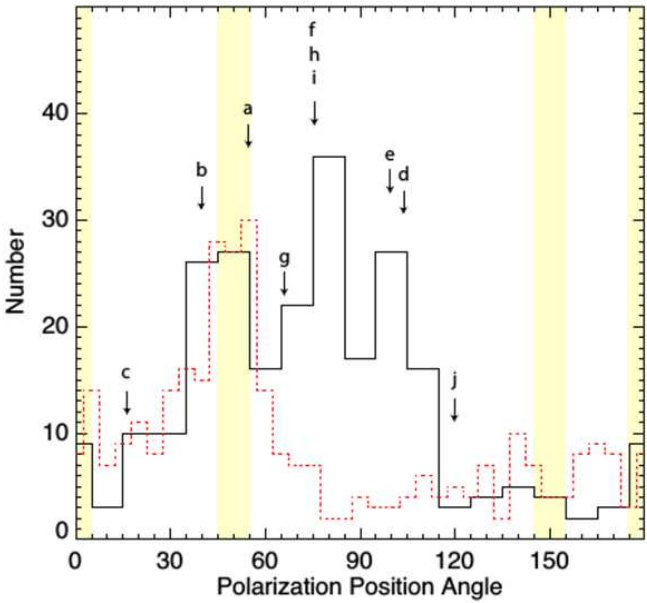}
\caption{
Histogram of polarization position angles for the 90$\arcdeg$-rotated submillimeter vectors 
with $I>0$, $P/ \delta P > 2$, and $\delta P < 4\%$ (black). 
Black line: 90$\arcdeg$ rotated submillimeter polarization vectors (this work). Note that a 10$\arcdeg$ bin is used.
Red dotted line: $H$-band  polarization position angles \citep{2015ApJS..220...17K}.
Black arrows indicate the magnetic field components from the submillimeter data. 
The labels a--j indicate the average magnetic field direction of the distinct magnetic field components a--j
projected on the plane of the sky.
Yellow colored regions indicate the major magnetic field directions suggested from the previous near-infrared data.
}
\label{fig:histo}
\end{figure}

Our new submillimeter polarimetry demonstrates that 
one of the main polarization position angles in Oph-A are approximately 50$\arcdeg$ (see Figures \ref{fig:submm}--\ref{fig:velocity}), 
and so are well aligned with the 50$\arcdeg$ magnetic field found in the near-infrared 
(\citealt{2015ApJS..220...17K}; see also Figure \ref{fig:submm90} for their comparison within the same field-of-view).
\cite{2015ApJS..220...17K} found that
the ``50$\arcdeg$ component'' is the dominant magnetic field component in the observed region; 
it can be seen as a distinct clump in the diagram plotting degree of polarization versus polarization angle 
(Figure 9 of \citealt{2015ApJS..220...17K}) 
and in the histogram of polarization position angles (Figure 10 of \citealt{2015ApJS..220...17K}). 
This component is seen in the northeast regions of $\rho$ Oph-A (and in $\rho$ Oph-B and $\rho$ Oph-E in a large scale, 
regions not covered in this work).
The ``0$\arcdeg$ component'' can be seen from $\rho$ Oph-A 
toward $\rho$ Oph-AC (located at the southeastern region of $\rho$ Oph-A, which is not shown in our submillimeter map; 
cf. \citealt{2015ApJS..220...17K}). 
In contrast, in Oph-A, both the 0$\arcdeg$ and the 50$\arcdeg$ components exist.

Figure \ref{fig:histo} shows the histogram of polarization position angles 
for the 90$\arcdeg$ rotated submillimeter polarization vectors 
as well as for the $H$-band  polarization position angles from \cite{2015ApJS..220...17K}. 
The distribution is relatively widespread, but 
if we refer to both the $H$-band polarization vector map (Figure 8 of \citealt{2015ApJS..220...17K}) and this histogram, 
we see several components, of which the components at
0\arcdeg and 50$\arcdeg$ are most clearly seen.
As shown in Figure \ref{fig:histo},
the distribution of the polarization position angles obtained from submillimeter polarimetry 
is in relatively good agreement with those obtained from near-infrared polarimetry for the 
0$\arcdeg$ and 50$\arcdeg$ components but not for the 150$\arcdeg$ component.
Note that 
since our submillimeter map covers a small part of the area covered by the near-infrared polarimetry survey
and we see much higher-column-density regions of $\rho$ Oph-A,
there is also some inconsistency between the distributions of submillimeter and near-infrared polarization angles.
Therefore, our results indicate that both submillimeter emission polarization and near-infrared dichroic polarization 
may trace the magnetic field structures associated with the $\rho$ Oph-A core region
at different spatial scales and at different region along the line-of-sight.

Previous observations have shown agreement between the magnetic field structures seen at various wavelengths 
such as near- and far-infrared, or submillimeter wavelengths 
(e.g., \citealp{1996ASPC...97..372T, 2007PASJ...59..467T, 2007PASJ...59..487K, 2011ApJ...741...35K}). 
Our new results are consistent with this behavior, although the greatest density regions can be traced only by submillimeter polarimetry.
The 50$\arcdeg$ component seen in the lower-density regions of the submillimeter map around the edge of the core
(the north-east filament, cf. Figure \ref{fig:OphA_stokesI}) is
consistent with the 50$\arcdeg$ component seen in the lower-density tracer of near-infrared polarization \citep{2015ApJS..220...17K},
giving us still further confidence in our observations.
Our data are also consistent with the recently released HAWC+ data taken by SOFIA \citep{2018AAS...23113004S}.
A combination of polarimetric observations over wavelengths and scales observed by instruments such as ALMA and by 8-m class optical/infrared telescopes will become more important in the future, to test the range of scales over which this behavior holds.

\section{SUMMARY}    \label{sec:summary}

In this paper, we present the first-look analysis for the $\rho$ Oph-A SCUBA-2/POL-2 continuum map
observed by the JCMT Gould Belt polarization survey at 850 $\mu$m.
The $\rho$ Oph molecular cloud complex is one of the nearest laboratories 
for examining active star-formation sites,
offering a wealth of objects to aid in a better understanding of the dominant physical processes present in the region. 
The SCUBA-2/POL-2 polarimeter is a very powerful instrument to trace the magnetic eld structure in star forming regions 
such as the  Oph molecular cloud complex.
The main results are as follows.

\begin{enumerate}
\item
We have identified at least 10 magnetic field components in the $\rho$ Oph-A core region,
whose position angles and degrees of polarization are distinct from each other. 
However, some of them can be part of a coherent structure. Our
polarimetric results are not only consistent with previous results in the bright core regions, 
but also reveal the fields in the outer regions for the first time.
These components represent the magnetic fields of the sub-cores identified as local continuum intensity peaks or distinct velocity structures within the Oph-A core; 
they show a large variation even within the small (approximately 0.2 pc) region observed.

\item
The dominant component of the magnetic field over $\rho$ Oph-A is the 50$\arcdeg$ component.
This direction is consistent with that inferred from the near-infrared polarimetry of the $\rho$ Oph cloud core.

\item
Although the average direction of the main component is approximately 50$\arcdeg$, 
the magnetic field tends to be locally perpendicular (approximately 100$\arcdeg$--110$\arcdeg$) to the arc-structure.
Between SM~1N and A-MM~6, the field direction is almost east-west, 
while the arc extends to the north-east or north-west, 
and the field direction extends toward LFAM~1 and GSS 30--1.
The perpendicularity between the core shape and the magnetic field direction 
may be important in understanding the origin and formation of this core.
Such perpendicularity is often seen in the densest parts of clouds and cloud cores.

\item
There are local structures besides the 50$\arcdeg$ component.
To the north of A-MM~6, the field direction is almost north-south, and
to the north of A-MM~7, the field direction is almost north-east, which is the same of the direction of the north-east filament.
Notable are the low degree of polarization near SM~1N 
and some deviation in field direction near VLA~1623 and its outflow region.

\item
Using the Davis-Chandrasekhar-Fermi method,
we roughly estimate the strengths of the magnetic field projected on the plane of the sky in several sub-core regions to be 
up to a few mG.

\item
We have found that the main large-scale core velocity gradient is approximately perpendicular to the inferred cloud magnetic field orientation.

\end{enumerate}

\acknowledgments

We thank the referee for thorough and insightful comments,
which improved the paper significantly.
The James Clerk Maxwell Telescope is operated by the 
East Asian Observatory on behalf of The National Astronomical 
Observatory of Japan, Academia Sinica Institute 
of Astronomy and Astrophysics, the Korea Astronomy 
and Space Science Institute, the National Astronomical 
Observatories of China and the Chinese Academy of Sciences 
(grant No.XDB09000000), with additional funding 
support from the Science and Technology Facilities 
Council of the United Kingdom and participating universities 
in the United Kingdom and Canada.
The James 
Clerk Maxwell Telescope has historically been operated 
by the Joint Astronomy Centre on behalf of the Science
and Technology Facilities Council of the United Kingdom,
the National Research Council of Canada and the 
Netherlands Organisation for Scientific Research.
Additional funds for the construction of SCUBA-2 and POL-2 
were provided by the Canada Foundation for Innovation. 
The data taken in this paper were observed under the project code M16AL004. 
Data analysis was in part carried out on the open use data analysis computer system 
at the Astronomy Data Center, ADC, of the National Astronomical Observatory of Japan.
J.K. 
was supported by MEXT KAKENHI grant number 16H07479
and the Astrobiology Center of NINS.
M.T. 
was supported by MEXT KAKENHI grant number 22000005. 
D.W.T. and K.P. 
acknowledge Science and Technology Facilities Council (STFC) support under grant numbers ST/K002023/1 and
ST/M000877/1. 
K.P. 
was an International Research Fellow of the Japan Society for the Promotion of Science.
K.P and S.P.L. 
acknowledge the support of the Ministry of
Science and Technology of Taiwan (Grant No. 106-2119-M-007 -021 -MY3).
M.K. 
was supported by Basic Science Research Program through the National Research Foundation of Korea (NRF) funded by the Ministry of
Science, ICT \& Future Planning (No. NRF-2015R1C1A1A01052160).
C.W.L. 
was supported by Basic Science Research Program though the National Research Foundation of Korea (NRF) funded by the Ministry of Education, Science, and Technology (NRF-2016R1A2B4012593).
W.K. 
was supported by Basic Science Research Program through the National Research Foundation of Korea (NRF-2016R1C1B2013642).
T.L. is supported by KASI fellowship and EACOA fellowship.
Team BISTRO-J is in part financially supported by 260 individuals.

\vspace{5mm}
\facility{James Clerk Maxwell Telescope (JCMT).}
\software{Starlink \citep{2014ASPC..485..391C}, 
	smurf (\citealt{2005ASPC..343...71B, 2013MNRAS.430.2545C})
	}

\clearpage
\startlongtable 
\begin{deluxetable*}{rccrrrrrc}
\tabletypesize{\tiny} 
\tablewidth{0pt}
\tablecaption{Submillimeter Polarimetry in the $\rho$ Ophiuchi Cloud Core}
\tablehead{
  \colhead{ID} 	&\multicolumn{2}{c}{Position} 				& \colhead{$I \pm \delta I$} & \colhead{$Q \pm  \delta Q$} & \colhead{$U \pm  \delta U$} 	& \colhead{$P \pm  \delta P$} 	& \colhead{$\theta \pm \delta \theta$} &  \colhead{Component }
\\\cline{2-3}
 \colhead{ }         	&\colhead {$\alpha_{\rm J2000}$} 		& \colhead {$\delta_{\rm J2000}$} & \colhead{ (mJy/beam) } & \colhead{  (mJy/beam) } & \colhead{  (mJy/beam) } & 
\colhead{(\%)}		& \colhead{(\arcdeg)}  & \colhead{ }
}
\startdata
1	&	16:26:33.4	&	-24:26:35.10	&	51.430	$\pm$	2.098	&	-0.342	$\pm$	1.619	&	4.101	$\pm$	1.608	&	7.37	$\pm$	3.14	&	47.4	$\pm$	11.3	&	j	\\
2	&	16:26:34.3	&	-24:26:23.10	&	58.272	$\pm$	2.122	&	-1.394	$\pm$	1.615	&	4.885	$\pm$	1.616	&	8.26	$\pm$	2.79	&	53.0	$\pm$	9.1	&	j	\\
3	&	16:26:32.5	&	-24:26:23.10	&	62.223	$\pm$	2.095	&	-2.642	$\pm$	1.595	&	3.811	$\pm$	1.592	&	7.00	$\pm$	2.57	&	62.4	$\pm$	9.8	&	j	\\
4	&	16:26:33.4	&	-24:26:23.10	&	67.742	$\pm$	2.103	&	2.010	$\pm$	1.603	&	6.867	$\pm$	1.597	&	10.30	$\pm$	2.38	&	36.8	$\pm$	6.4	&	j	\\
5	&	16:26:33.4	&	-24:26:11.10	&	69.851	$\pm$	2.038	&	1.644	$\pm$	1.576	&	10.126	$\pm$	1.563	&	14.51	$\pm$	2.28	&	40.4	$\pm$	4.4	&	j	\\
6	&	16:26:32.5	&	-24:26:11.10	&	74.294	$\pm$	2.043	&	1.045	$\pm$	1.552	&	4.014	$\pm$	1.556	&	5.18	$\pm$	2.10	&	37.7	$\pm$	10.7	&	j	\\
7	&	16:26:32.5	&	-24:25:59.10	&	50.015	$\pm$	2.039	&	2.705	$\pm$	1.519	&	4.317	$\pm$	1.510	&	9.73	$\pm$	3.05	&	29.0	$\pm$	8.5	&	j	\\
8	&	16:26:32.5	&	-24:25:35.10	&	39.887	$\pm$	2.062	&	10.629	$\pm$	1.467	&	1.343	$\pm$	1.482	&	26.61	$\pm$	3.93	&	3.6	$\pm$	4.0	&	j	\\
9	&	16:26:34.3	&	-24:25:23.10	&	61.159	$\pm$	2.082	&	2.735	$\pm$	1.496	&	4.664	$\pm$	1.506	&	8.49	$\pm$	2.48	&	29.8	$\pm$	7.9	&	j	\\
10	&	16:26:33.4	&	-24:25:23.10	&	106.577	$\pm$	2.065	&	5.813	$\pm$	1.473	&	4.804	$\pm$	1.484	&	6.94	$\pm$	1.39	&	19.8	$\pm$	5.6	&	j	\\
11	&	16:26:32.5	&	-24:25:23.10	&	136.809	$\pm$	2.039	&	9.366	$\pm$	1.450	&	5.233	$\pm$	1.458	&	7.77	$\pm$	1.07	&	14.6	$\pm$	3.9	&	j	\\
12	&	16:26:30.8	&	-24:25:23.10	&	140.545	$\pm$	2.035	&	7.720	$\pm$	1.442	&	5.518	$\pm$	1.452	&	6.67	$\pm$	1.03	&	17.8	$\pm$	4.4	&	j	\\
13	&	16:26:31.6	&	-24:25:23.10	&	158.406	$\pm$	2.045	&	10.415	$\pm$	1.443	&	3.233	$\pm$	1.449	&	6.82	$\pm$	0.92	&	8.6	$\pm$	3.8	&	j	\\
14	&	16:26:18.5	&	-24:25:23.08	&	60.328	$\pm$	2.255	&	-2.298	$\pm$	1.482	&	4.702	$\pm$	1.505	&	8.31	$\pm$	2.51	&	58.0	$\pm$	8.1	&	$\cdots$	\\
15	&	16:26:29.0	&	-24:25:11.10	&	72.854	$\pm$	2.037	&	5.357	$\pm$	1.399	&	-3.862	$\pm$	1.396	&	8.86	$\pm$	1.94	&	-17.9	$\pm$	6.1	&	$\cdots$	\\
16	&	16:26:34.3	&	-24:25:11.10	&	163.985	$\pm$	2.057	&	-2.271	$\pm$	1.473	&	5.622	$\pm$	1.495	&	3.58	$\pm$	0.91	&	56.0	$\pm$	7.0	&	i	\\
17	&	16:26:29.9	&	-24:25:11.10	&	179.956	$\pm$	2.042	&	1.486	$\pm$	1.404	&	-3.252	$\pm$	1.405	&	1.83	$\pm$	0.78	&	-32.7	$\pm$	11.3	&	i	\\
18	&	16:26:33.4	&	-24:25:11.10	&	243.825	$\pm$	2.065	&	5.194	$\pm$	1.455	&	0.295	$\pm$	1.469	&	2.05	$\pm$	0.60	&	1.6	$\pm$	8.1	&	i	\\
19	&	16:26:32.5	&	-24:25:11.10	&	263.398	$\pm$	2.060	&	11.865	$\pm$	1.441	&	0.168	$\pm$	1.450	&	4.47	$\pm$	0.55	&	0.4	$\pm$	3.5	&	i	\\
20	&	16:26:31.6	&	-24:25:11.10	&	285.549	$\pm$	2.022	&	15.906	$\pm$	1.409	&	2.776	$\pm$	1.435	&	5.63	$\pm$	0.50	&	5.0	$\pm$	2.5	&	i	\\
21	&	16:26:30.8	&	-24:25:11.10	&	296.876	$\pm$	2.016	&	16.204	$\pm$	1.409	&	1.031	$\pm$	1.408	&	5.45	$\pm$	0.48	&	1.8	$\pm$	2.5	&	i	\\
22	&	16:26:36.0	&	-24:25:11.09	&	52.454	$\pm$	2.123	&	2.267	$\pm$	1.542	&	4.078	$\pm$	1.549	&	8.39	$\pm$	2.97	&	30.5	$\pm$	9.5	&	$\cdots$	\\
23	&	16:26:35.2	&	-24:25:11.09	&	64.541	$\pm$	2.038	&	0.240	$\pm$	1.505	&	4.359	$\pm$	1.504	&	6.35	$\pm$	2.34	&	43.4	$\pm$	9.9	&	$\cdots$	\\
24	&	16:26:19.3	&	-24:25:11.08	&	57.627	$\pm$	2.173	&	-1.546	$\pm$	1.457	&	-2.992	$\pm$	1.477	&	5.26	$\pm$	2.57	&	-58.7	$\pm$	12.4	&	$\cdots$	\\
25	&	16:26:15.8	&	-24:25:11.06	&	48.901	$\pm$	2.371	&	5.178	$\pm$	1.543	&	-0.781	$\pm$	1.566	&	10.23	$\pm$	3.20	&	-4.3	$\pm$	8.6	&	$\cdots$	\\
26	&	16:26:28.1	&	-24:24:59.10	&	79.582	$\pm$	2.088	&	2.094	$\pm$	1.378	&	-4.110	$\pm$	1.383	&	5.53	$\pm$	1.74	&	-31.5	$\pm$	8.6	&	$\cdots$	\\
27	&	16:26:29.0	&	-24:24:59.10	&	209.209	$\pm$	2.086	&	7.624	$\pm$	1.374	&	-7.347	$\pm$	1.388	&	5.02	$\pm$	0.66	&	-22.0	$\pm$	3.7	&	$\cdots$	\\
28	&	16:26:34.3	&	-24:24:59.10	&	275.086	$\pm$	2.023	&	-3.445	$\pm$	1.464	&	-0.868	$\pm$	1.472	&	1.18	$\pm$	0.53	&	-82.9	$\pm$	11.9	&	i	\\
29	&	16:26:33.4	&	-24:24:59.10	&	407.349	$\pm$	2.051	&	7.166	$\pm$	1.441	&	2.139	$\pm$	1.454	&	1.80	$\pm$	0.35	&	8.3	$\pm$	5.6	&	i	\\
30	&	16:26:29.9	&	-24:24:59.10	&	421.360	$\pm$	2.050	&	9.406	$\pm$	1.387	&	-10.463	$\pm$	1.378	&	3.32	$\pm$	0.33	&	-24.0	$\pm$	2.8	&	i	\\
31	&	16:26:30.8	&	-24:24:59.10	&	472.817	$\pm$	2.054	&	19.633	$\pm$	1.387	&	-6.862	$\pm$	1.412	&	4.39	$\pm$	0.29	&	-9.6	$\pm$	1.9	&	i	\\
32	&	16:26:32.5	&	-24:24:59.10	&	506.147	$\pm$	2.067	&	14.517	$\pm$	1.422	&	-3.031	$\pm$	1.424	&	2.92	$\pm$	0.28	&	-5.9	$\pm$	2.8	&	i	\\
33	&	16:26:31.6	&	-24:24:59.10	&	535.088	$\pm$	2.043	&	20.017	$\pm$	1.408	&	-4.015	$\pm$	1.409	&	3.81	$\pm$	0.26	&	-5.7	$\pm$	2.0	&	i	\\
34	&	16:26:22.9	&	-24:24:59.09	&	36.637	$\pm$	2.072	&	4.997	$\pm$	1.386	&	-0.534	$\pm$	1.386	&	13.19	$\pm$	3.86	&	-3.1	$\pm$	7.9	&	h	\\
35	&	16:26:36.9	&	-24:24:59.09	&	48.555	$\pm$	2.180	&	3.892	$\pm$	1.580	&	0.981	$\pm$	1.573	&	7.60	$\pm$	3.27	&	7.1	$\pm$	11.2	&	$\cdots$	\\
36	&	16:26:36.0	&	-24:24:59.09	&	105.173	$\pm$	2.056	&	4.079	$\pm$	1.527	&	-1.656	$\pm$	1.529	&	3.93	$\pm$	1.45	&	-11.0	$\pm$	9.9	&	$\cdots$	\\
37	&	16:26:14.1	&	-24:24:59.05	&	56.258	$\pm$	2.532	&	3.874	$\pm$	1.604	&	3.473	$\pm$	1.615	&	8.80	$\pm$	2.89	&	20.9	$\pm$	8.9	&	$\cdots$	\\
38	&	16:26:25.5	&	-24:24:47.10	&	128.626	$\pm$	2.111	&	2.576	$\pm$	1.331	&	-3.453	$\pm$	1.347	&	3.18	$\pm$	1.04	&	-26.6	$\pm$	8.9	&	h	\\
39	&	16:26:27.3	&	-24:24:47.10	&	315.284	$\pm$	2.174	&	6.195	$\pm$	1.357	&	-1.949	$\pm$	1.352	&	2.01	$\pm$	0.43	&	-8.7	$\pm$	6.0	&	$\cdots$	\\
40	&	16:26:28.1	&	-24:24:47.10	&	322.238	$\pm$	2.094	&	3.581	$\pm$	1.354	&	-9.384	$\pm$	1.351	&	3.09	$\pm$	0.42	&	-34.6	$\pm$	3.9	&	$\cdots$	\\
41	&	16:26:33.4	&	-24:24:47.10	&	448.077	$\pm$	2.046	&	6.608	$\pm$	1.425	&	-0.974	$\pm$	1.431	&	1.46	$\pm$	0.32	&	-4.2	$\pm$	6.1	&	i	\\
42	&	16:26:29.0	&	-24:24:47.10	&	505.237	$\pm$	2.110	&	9.237	$\pm$	1.360	&	-16.107	$\pm$	1.369	&	3.67	$\pm$	0.27	&	-30.1	$\pm$	2.1	&	$\cdots$	\\
43	&	16:26:32.5	&	-24:24:47.10	&	608.888	$\pm$	2.048	&	12.679	$\pm$	1.403	&	-2.942	$\pm$	1.413	&	2.13	$\pm$	0.23	&	-6.5	$\pm$	3.1	&	i	\\
44	&	16:26:29.9	&	-24:24:47.10	&	781.876	$\pm$	2.061	&	20.867	$\pm$	1.368	&	-17.607	$\pm$	1.377	&	3.49	$\pm$	0.18	&	-20.1	$\pm$	1.4	&	i	\\
45	&	16:26:31.6	&	-24:24:47.10	&	834.935	$\pm$	2.056	&	24.301	$\pm$	1.389	&	-9.222	$\pm$	1.397	&	3.11	$\pm$	0.17	&	-10.4	$\pm$	1.5	&	i	\\
46	&	16:26:30.8	&	-24:24:47.10	&	834.771	$\pm$	2.034	&	21.231	$\pm$	1.376	&	-13.423	$\pm$	1.377	&	3.00	$\pm$	0.17	&	-16.2	$\pm$	1.6	&	i	\\
47	&	16:26:22.9	&	-24:24:47.09	&	38.256	$\pm$	2.075	&	2.747	$\pm$	1.353	&	-1.657	$\pm$	1.371	&	7.60	$\pm$	3.58	&	-15.6	$\pm$	12.2	&	h	\\
48	&	16:26:23.7	&	-24:24:47.09	&	46.716	$\pm$	2.083	&	1.630	$\pm$	1.347	&	-3.585	$\pm$	1.355	&	7.92	$\pm$	2.92	&	-32.8	$\pm$	9.8	&	h	\\
49	&	16:26:33.4	&	-24:24:35.10	&	375.056	$\pm$	2.013	&	-1.943	$\pm$	1.399	&	-5.926	$\pm$	1.412	&	1.62	$\pm$	0.38	&	-54.1	$\pm$	6.4	&	i	\\
50	&	16:26:32.5	&	-24:24:35.10	&	550.300	$\pm$	2.058	&	3.323	$\pm$	1.387	&	-6.657	$\pm$	1.393	&	1.33	$\pm$	0.25	&	-31.7	$\pm$	5.3	&	i	\\
51	&	16:26:25.5	&	-24:24:35.10	&	553.396	$\pm$	2.136	&	5.956	$\pm$	1.308	&	-5.067	$\pm$	1.317	&	1.39	$\pm$	0.24	&	-20.2	$\pm$	4.8	&	h	\\
52	&	16:26:31.6	&	-24:24:35.10	&	930.155	$\pm$	2.048	&	16.612	$\pm$	1.368	&	-14.349	$\pm$	1.370	&	2.36	$\pm$	0.15	&	-20.4	$\pm$	1.8	&	i	\\
53	&	16:26:28.1	&	-24:24:35.10	&	1003.480	$\pm$	2.149	&	9.105	$\pm$	1.330	&	-12.010	$\pm$	1.332	&	1.50	$\pm$	0.13	&	-26.4	$\pm$	2.5	&	a	\\
54	&	16:26:29.0	&	-24:24:35.10	&	1119.110	$\pm$	2.061	&	17.792	$\pm$	1.332	&	-14.137	$\pm$	1.336	&	2.03	$\pm$	0.12	&	-19.2	$\pm$	1.7	&	a	\\
55	&	16:26:30.8	&	-24:24:35.10	&	1174.260	$\pm$	2.021	&	20.439	$\pm$	1.352	&	-20.801	$\pm$	1.362	&	2.48	$\pm$	0.12	&	-22.8	$\pm$	1.3	&	i	\\
56	&	16:26:29.9	&	-24:24:35.10	&	1282.560	$\pm$	2.076	&	19.135	$\pm$	1.338	&	-24.433	$\pm$	1.349	&	2.42	$\pm$	0.10	&	-26.0	$\pm$	1.2	&	i	\\
57	&	16:26:27.3	&	-24:24:35.10	&	1371.780	$\pm$	2.362	&	7.661	$\pm$	1.329	&	-11.563	$\pm$	1.319	&	1.01	$\pm$	0.10	&	-28.2	$\pm$	2.7	&	a	\\
58	&	16:26:26.4	&	-24:24:35.10	&	1887.860	$\pm$	2.594	&	3.423	$\pm$	1.319	&	-17.786	$\pm$	1.322	&	0.96	$\pm$	0.07	&	-39.6	$\pm$	2.1	&	a	\\
59	&	16:26:22.9	&	-24:24:35.09	&	47.454	$\pm$	2.053	&	3.372	$\pm$	1.335	&	-0.894	$\pm$	1.355	&	6.79	$\pm$	2.83	&	-7.4	$\pm$	11.1	&	h	\\
60	&	16:26:35.2	&	-24:24:35.09	&	96.748	$\pm$	2.068	&	-3.614	$\pm$	1.467	&	1.021	$\pm$	1.481	&	3.57	$\pm$	1.52	&	82.1	$\pm$	11.3	&	$\cdots$	\\
61	&	16:26:23.7	&	-24:24:35.09	&	168.924	$\pm$	2.043	&	3.591	$\pm$	1.313	&	0.056	$\pm$	1.332	&	1.98	$\pm$	0.78	&	0.4	$\pm$	10.6	&	h	\\
62	&	16:26:34.3	&	-24:24:23.10	&	162.147	$\pm$	2.032	&	1.783	$\pm$	1.437	&	-7.161	$\pm$	1.433	&	4.46	$\pm$	0.89	&	-38.0	$\pm$	5.6	&	i	\\
63	&	16:26:33.4	&	-24:24:23.10	&	333.190	$\pm$	2.020	&	-2.890	$\pm$	1.380	&	-10.016	$\pm$	1.409	&	3.10	$\pm$	0.42	&	-53.0	$\pm$	3.8	&	i	\\
64	&	16:26:32.5	&	-24:24:23.10	&	534.381	$\pm$	1.993	&	1.523	$\pm$	1.365	&	-8.430	$\pm$	1.379	&	1.58	$\pm$	0.26	&	-39.9	$\pm$	4.6	&	i	\\
65	&	16:26:25.5	&	-24:24:23.10	&	778.337	$\pm$	2.115	&	-0.120	$\pm$	1.295	&	-5.106	$\pm$	1.291	&	0.63	$\pm$	0.17	&	-45.7	$\pm$	7.3	&	$\cdots$	\\
66	&	16:26:31.6	&	-24:24:23.10	&	816.746	$\pm$	2.022	&	4.701	$\pm$	1.351	&	-20.827	$\pm$	1.353	&	2.61	$\pm$	0.17	&	-38.6	$\pm$	1.8	&	i	\\
67	&	16:26:30.8	&	-24:24:23.10	&	1158.130	$\pm$	1.998	&	11.575	$\pm$	1.341	&	-27.026	$\pm$	1.337	&	2.54	$\pm$	0.12	&	-33.4	$\pm$	1.3	&	i	\\
68	&	16:26:29.9	&	-24:24:23.10	&	1493.050	$\pm$	2.017	&	12.956	$\pm$	1.320	&	-31.557	$\pm$	1.326	&	2.28	$\pm$	0.09	&	-33.8	$\pm$	1.1	&	i	\\
69	&	16:26:29.0	&	-24:24:23.10	&	1660.180	$\pm$	2.071	&	8.695	$\pm$	1.314	&	-23.456	$\pm$	1.304	&	1.50	$\pm$	0.08	&	-34.8	$\pm$	1.5	&	a	\\
70	&	16:26:28.1	&	-24:24:23.10	&	1964.770	$\pm$	2.153	&	3.659	$\pm$	1.314	&	-30.492	$\pm$	1.312	&	1.56	$\pm$	0.07	&	-41.6	$\pm$	1.2	&	a	\\
71	&	16:26:26.4	&	-24:24:23.10	&	2416.880	$\pm$	2.535	&	-4.410	$\pm$	1.289	&	-25.770	$\pm$	1.303	&	1.08	$\pm$	0.05	&	-49.9	$\pm$	1.4	&	a	\\
72	&	16:26:27.3	&	-24:24:23.10	&	2643.620	$\pm$	2.315	&	4.500	$\pm$	1.290	&	-42.859	$\pm$	1.297	&	1.63	$\pm$	0.05	&	-42.0	$\pm$	0.9	&	a	\\
73	&	16:26:23.7	&	-24:24:23.09	&	136.669	$\pm$	2.065	&	-3.840	$\pm$	1.302	&	3.425	$\pm$	1.307	&	3.64	$\pm$	0.96	&	69.1	$\pm$	7.3	&	$\cdots$	\\
74	&	16:26:34.3	&	-24:24:11.10	&	86.585	$\pm$	2.059	&	4.128	$\pm$	1.419	&	-6.984	$\pm$	1.433	&	9.22	$\pm$	1.67	&	-29.7	$\pm$	5.0	&	g	\\
75	&	16:26:33.4	&	-24:24:11.10	&	200.755	$\pm$	2.019	&	1.662	$\pm$	1.381	&	-9.789	$\pm$	1.392	&	4.90	$\pm$	0.70	&	-40.2	$\pm$	4.0	&	g	\\
76	&	16:26:32.5	&	-24:24:11.10	&	354.117	$\pm$	1.999	&	-0.931	$\pm$	1.355	&	-17.182	$\pm$	1.359	&	4.84	$\pm$	0.38	&	-46.6	$\pm$	2.3	&	g	\\
77	&	16:26:31.6	&	-24:24:11.10	&	454.205	$\pm$	1.989	&	3.053	$\pm$	1.323	&	-22.731	$\pm$	1.347	&	5.04	$\pm$	0.30	&	-41.2	$\pm$	1.7	&	g	\\
78	&	16:26:25.5	&	-24:24:11.10	&	473.293	$\pm$	2.094	&	1.082	$\pm$	1.276	&	-8.387	$\pm$	1.288	&	1.77	$\pm$	0.27	&	-41.3	$\pm$	4.3	&	f	\\
79	&	16:26:30.8	&	-24:24:11.10	&	627.260	$\pm$	2.010	&	3.335	$\pm$	1.322	&	-28.680	$\pm$	1.329	&	4.60	$\pm$	0.21	&	-41.7	$\pm$	1.3	&	g	\\
80	&	16:26:29.9	&	-24:24:11.10	&	953.887	$\pm$	2.014	&	9.991	$\pm$	1.303	&	-26.395	$\pm$	1.314	&	2.96	$\pm$	0.14	&	-34.6	$\pm$	1.3	&	g	\\
81	&	16:26:26.4	&	-24:24:11.10	&	1417.450	$\pm$	2.182	&	11.137	$\pm$	1.272	&	-24.760	$\pm$	1.281	&	1.91	$\pm$	0.09	&	-32.9	$\pm$	1.3	&	a	\\
82	&	16:26:29.0	&	-24:24:11.10	&	1576.900	$\pm$	2.115	&	1.175	$\pm$	1.300	&	-26.510	$\pm$	1.300	&	1.68	$\pm$	0.08	&	-43.7	$\pm$	1.4	&	a	\\
83	&	16:26:27.3	&	-24:24:11.10	&	2618.600	$\pm$	2.216	&	6.129	$\pm$	1.275	&	-57.370	$\pm$	1.284	&	2.20	$\pm$	0.05	&	-42.0	$\pm$	0.6	&	a	\\
84	&	16:26:28.1	&	-24:24:11.10	&	2697.520	$\pm$	2.215	&	-6.913	$\pm$	1.285	&	-40.965	$\pm$	1.300	&	1.54	$\pm$	0.05	&	-49.8	$\pm$	0.9	&	a	\\
85	&	16:26:22.9	&	-24:24:11.09	&	50.474	$\pm$	2.353	&	-8.634	$\pm$	1.307	&	3.476	$\pm$	1.310	&	18.26	$\pm$	2.73	&	79.0	$\pm$	4.0	&	f	\\
86	&	16:26:23.7	&	-24:24:11.09	&	127.343	$\pm$	2.063	&	-6.138	$\pm$	1.283	&	1.183	$\pm$	1.297	&	4.80	$\pm$	1.01	&	84.5	$\pm$	5.9	&	f	\\
87	&	16:26:33.4	&	-24:23:59.10	&	68.829	$\pm$	2.028	&	-1.088	$\pm$	1.388	&	-3.856	$\pm$	1.393	&	5.46	$\pm$	2.03	&	-52.9	$\pm$	9.9	&	g	\\
88	&	16:26:32.5	&	-24:23:59.10	&	144.438	$\pm$	1.971	&	3.840	$\pm$	1.353	&	-11.073	$\pm$	1.367	&	8.06	$\pm$	0.95	&	-35.4	$\pm$	3.3	&	g	\\
89	&	16:26:31.6	&	-24:23:59.10	&	201.312	$\pm$	1.996	&	7.689	$\pm$	1.324	&	-18.711	$\pm$	1.332	&	10.03	$\pm$	0.67	&	-33.8	$\pm$	1.9	&	g	\\
90	&	16:26:25.5	&	-24:23:59.10	&	254.610	$\pm$	2.083	&	9.190	$\pm$	1.261	&	-3.073	$\pm$	1.279	&	3.77	$\pm$	0.50	&	-9.2	$\pm$	3.8	&	f	\\
91	&	16:26:30.8	&	-24:23:59.10	&	300.531	$\pm$	2.007	&	8.665	$\pm$	1.317	&	-18.943	$\pm$	1.315	&	6.92	$\pm$	0.44	&	-32.7	$\pm$	1.8	&	g	\\
92	&	16:26:29.9	&	-24:23:59.10	&	524.082	$\pm$	2.014	&	10.789	$\pm$	1.291	&	-21.277	$\pm$	1.305	&	4.55	$\pm$	0.25	&	-31.6	$\pm$	1.6	&	g	\\
93	&	16:26:29.0	&	-24:23:59.10	&	1115.710	$\pm$	2.140	&	17.291	$\pm$	1.284	&	-27.305	$\pm$	1.295	&	2.89	$\pm$	0.12	&	-28.8	$\pm$	1.1	&	a	\\
94	&	16:26:26.4	&	-24:23:59.10	&	1292.280	$\pm$	2.185	&	26.583	$\pm$	1.273	&	-17.224	$\pm$	1.278	&	2.45	$\pm$	0.10	&	-16.5	$\pm$	1.2	&	a	\\
95	&	16:26:27.3	&	-24:23:59.10	&	3352.910	$\pm$	2.478	&	23.719	$\pm$	1.283	&	-75.009	$\pm$	1.285	&	2.35	$\pm$	0.04	&	-36.2	$\pm$	0.5	&	a	\\
96	&	16:26:28.1	&	-24:23:59.10	&	3402.040	$\pm$	2.498	&	11.316	$\pm$	1.285	&	-63.491	$\pm$	1.298	&	1.90	$\pm$	0.04	&	-39.9	$\pm$	0.6	&	a	\\
97	&	16:26:22.9	&	-24:23:59.09	&	81.824	$\pm$	2.074	&	-13.510	$\pm$	1.310	&	-1.131	$\pm$	1.321	&	16.49	$\pm$	1.66	&	-87.6	$\pm$	2.8	&	f	\\
98	&	16:26:23.7	&	-24:23:59.09	&	82.906	$\pm$	2.072	&	-4.200	$\pm$	1.289	&	1.345	$\pm$	1.299	&	5.09	$\pm$	1.56	&	81.1	$\pm$	8.4	&	f	\\
99	&	16:26:32.5	&	-24:23:47.10	&	43.684	$\pm$	1.993	&	6.401	$\pm$	1.355	&	-2.398	$\pm$	1.360	&	15.34	$\pm$	3.18	&	-10.3	$\pm$	5.7	&	g	\\
100	&	16:26:31.6	&	-24:23:47.10	&	67.312	$\pm$	1.970	&	9.691	$\pm$	1.331	&	-7.843	$\pm$	1.335	&	18.42	$\pm$	2.05	&	-19.5	$\pm$	3.1	&	g	\\
101	&	16:26:30.8	&	-24:23:47.10	&	143.749	$\pm$	1.981	&	14.878	$\pm$	1.313	&	-9.170	$\pm$	1.314	&	12.12	$\pm$	0.93	&	-15.8	$\pm$	2.2	&	g	\\
102	&	16:26:25.5	&	-24:23:47.10	&	237.806	$\pm$	2.070	&	6.513	$\pm$	1.263	&	1.656	$\pm$	1.276	&	2.78	$\pm$	0.53	&	7.1	$\pm$	5.4	&	f	\\
103	&	16:26:29.9	&	-24:23:47.10	&	304.683	$\pm$	2.024	&	20.928	$\pm$	1.292	&	-8.904	$\pm$	1.301	&	7.45	$\pm$	0.43	&	-11.5	$\pm$	1.6	&	g	\\
104	&	16:26:29.0	&	-24:23:47.10	&	761.543	$\pm$	2.091	&	18.376	$\pm$	1.273	&	-16.700	$\pm$	1.291	&	3.26	$\pm$	0.17	&	-21.1	$\pm$	1.5	&	a	\\
105	&	16:26:26.4	&	-24:23:47.10	&	1388.390	$\pm$	2.214	&	18.759	$\pm$	1.263	&	-12.531	$\pm$	1.278	&	1.62	$\pm$	0.09	&	-16.9	$\pm$	1.6	&	a	\\
106	&	16:26:28.1	&	-24:23:47.10	&	2694.310	$\pm$	2.563	&	4.042	$\pm$	1.289	&	-39.784	$\pm$	1.301	&	1.48	$\pm$	0.05	&	-42.1	$\pm$	0.9	&	a	\\
107	&	16:26:27.3	&	-24:23:47.10	&	3342.890	$\pm$	2.437	&	3.979	$\pm$	1.275	&	-65.795	$\pm$	1.288	&	1.97	$\pm$	0.04	&	-43.3	$\pm$	0.6	&	a	\\
108	&	16:26:21.1	&	-24:23:47.09	&	75.114	$\pm$	2.166	&	0.361	$\pm$	1.367	&	-3.412	$\pm$	1.369	&	4.19	$\pm$	1.83	&	-42.0	$\pm$	11.4	&	f	\\
109	&	16:26:22.0	&	-24:23:47.09	&	90.150	$\pm$	2.150	&	-4.675	$\pm$	1.347	&	-0.223	$\pm$	1.340	&	4.97	$\pm$	1.50	&	-88.6	$\pm$	8.2	&	f	\\
110	&	16:26:22.9	&	-24:23:47.09	&	104.711	$\pm$	2.086	&	-8.144	$\pm$	1.315	&	-2.148	$\pm$	1.329	&	7.94	$\pm$	1.27	&	-82.6	$\pm$	4.5	&	f	\\
111	&	16:26:17.6	&	-24:23:47.07	&	81.837	$\pm$	2.416	&	-0.953	$\pm$	1.480	&	4.201	$\pm$	1.473	&	4.95	$\pm$	1.81	&	51.4	$\pm$	9.8	&	$\cdots$	\\
112	&	16:26:24.6	&	-24:23:35.10	&	34.165	$\pm$	2.103	&	5.500	$\pm$	1.287	&	-0.679	$\pm$	1.299	&	15.78	$\pm$	3.90	&	-3.5	$\pm$	6.7	&	f	\\
113	&	16:26:30.8	&	-24:23:35.10	&	98.238	$\pm$	2.043	&	14.844	$\pm$	1.313	&	-4.484	$\pm$	1.322	&	15.73	$\pm$	1.38	&	-8.4	$\pm$	2.4	&	g	\\
114	&	16:26:29.9	&	-24:23:35.10	&	283.551	$\pm$	2.024	&	18.141	$\pm$	1.301	&	-3.264	$\pm$	1.311	&	6.48	$\pm$	0.46	&	-5.1	$\pm$	2.0	&	g	\\
115	&	16:26:25.5	&	-24:23:35.10	&	303.698	$\pm$	2.099	&	3.975	$\pm$	1.272	&	3.232	$\pm$	1.276	&	1.63	$\pm$	0.42	&	19.6	$\pm$	7.1	&	f	\\
116	&	16:26:29.0	&	-24:23:35.10	&	676.103	$\pm$	2.073	&	19.248	$\pm$	1.289	&	-5.387	$\pm$	1.297	&	2.95	$\pm$	0.19	&	-7.8	$\pm$	1.9	&	a	\\
117	&	16:26:26.4	&	-24:23:35.10	&	1357.650	$\pm$	2.208	&	3.294	$\pm$	1.273	&	-0.445	$\pm$	1.283	&	0.23	$\pm$	0.09	&	-3.8	$\pm$	11.1	&	a	\\
118	&	16:26:28.1	&	-24:23:35.10	&	2104.370	$\pm$	2.360	&	13.236	$\pm$	1.290	&	-4.725	$\pm$	1.297	&	0.67	$\pm$	0.06	&	-9.8	$\pm$	2.6	&	a	\\
119	&	16:26:27.3	&	-24:23:35.10	&	2889.410	$\pm$	2.241	&	0.268	$\pm$	1.270	&	-14.623	$\pm$	1.284	&	0.50	$\pm$	0.04	&	-44.5	$\pm$	2.5	&	a	\\
120	&	16:26:22.9	&	-24:23:35.09	&	133.382	$\pm$	2.143	&	-3.114	$\pm$	1.325	&	-0.235	$\pm$	1.336	&	2.12	$\pm$	0.99	&	-87.8	$\pm$	12.3	&	f	\\
121	&	16:26:21.1	&	-24:23:35.09	&	141.695	$\pm$	2.191	&	-1.824	$\pm$	1.382	&	-7.951	$\pm$	1.390	&	5.67	$\pm$	0.98	&	-51.5	$\pm$	4.9	&	f	\\
122	&	16:26:22.0	&	-24:23:35.09	&	176.072	$\pm$	2.160	&	-5.984	$\pm$	1.356	&	-5.542	$\pm$	1.369	&	4.57	$\pm$	0.78	&	-68.6	$\pm$	4.8	&	f	\\
123	&	16:26:20.2	&	-24:23:35.08	&	68.531	$\pm$	2.214	&	-3.665	$\pm$	1.414	&	-5.246	$\pm$	1.406	&	9.11	$\pm$	2.08	&	-62.5	$\pm$	6.3	&	f	\\
124	&	16:26:17.6	&	-24:23:35.07	&	45.658	$\pm$	2.459	&	-3.891	$\pm$	1.486	&	3.613	$\pm$	1.498	&	11.16	$\pm$	3.33	&	68.6	$\pm$	8.1	&	$\cdots$	\\
125	&	16:26:24.6	&	-24:23:23.10	&	72.227	$\pm$	2.084	&	6.803	$\pm$	1.299	&	-1.181	$\pm$	1.310	&	9.39	$\pm$	1.82	&	-4.9	$\pm$	5.4	&	e	\\
126	&	16:26:30.8	&	-24:23:23.10	&	85.309	$\pm$	2.073	&	10.189	$\pm$	1.334	&	1.491	$\pm$	1.335	&	11.97	$\pm$	1.59	&	4.2	$\pm$	3.7	&	g	\\
127	&	16:26:29.9	&	-24:23:23.10	&	310.310	$\pm$	2.079	&	16.010	$\pm$	1.312	&	-2.907	$\pm$	1.329	&	5.23	$\pm$	0.42	&	-5.1	$\pm$	2.3	&	g	\\
128	&	16:26:25.5	&	-24:23:23.10	&	321.875	$\pm$	2.090	&	4.902	$\pm$	1.290	&	-0.923	$\pm$	1.295	&	1.50	$\pm$	0.40	&	-5.3	$\pm$	7.4	&	e	\\
129	&	16:26:29.0	&	-24:23:23.10	&	701.789	$\pm$	2.107	&	18.357	$\pm$	1.302	&	-0.381	$\pm$	1.323	&	2.61	$\pm$	0.19	&	-0.6	$\pm$	2.1	&	e	\\
130	&	16:26:26.4	&	-24:23:23.10	&	1168.270	$\pm$	2.178	&	2.500	$\pm$	1.283	&	3.359	$\pm$	1.291	&	0.34	$\pm$	0.11	&	26.7	$\pm$	8.8	&	e	\\
131	&	16:26:28.1	&	-24:23:23.10	&	1718.330	$\pm$	2.270	&	19.074	$\pm$	1.298	&	8.152	$\pm$	1.310	&	1.20	$\pm$	0.08	&	11.6	$\pm$	1.8	&	e	\\
132	&	16:26:27.3	&	-24:23:23.10	&	2362.890	$\pm$	2.281	&	10.365	$\pm$	1.290	&	12.325	$\pm$	1.292	&	0.68	$\pm$	0.05	&	25.0	$\pm$	2.3	&	e	\\
133	&	16:26:22.9	&	-24:23:23.09	&	97.956	$\pm$	2.203	&	4.341	$\pm$	1.339	&	-2.653	$\pm$	1.346	&	5.01	$\pm$	1.37	&	-15.7	$\pm$	7.6	&	f	\\
134	&	16:26:21.1	&	-24:23:23.09	&	122.827	$\pm$	2.216	&	1.495	$\pm$	1.410	&	-7.600	$\pm$	1.403	&	6.20	$\pm$	1.15	&	-39.4	$\pm$	5.2	&	f	\\
135	&	16:26:22.0	&	-24:23:23.09	&	162.363	$\pm$	2.222	&	-0.593	$\pm$	1.368	&	-4.353	$\pm$	1.383	&	2.57	$\pm$	0.85	&	-48.9	$\pm$	8.9	&	f	\\
136	&	16:26:19.3	&	-24:23:23.08	&	40.275	$\pm$	2.341	&	-6.088	$\pm$	1.449	&	-3.108	$\pm$	1.464	&	16.59	$\pm$	3.74	&	-76.5	$\pm$	6.1	&	$\cdots$	\\
137	&	16:26:20.2	&	-24:23:23.08	&	66.519	$\pm$	2.301	&	-1.848	$\pm$	1.412	&	-5.787	$\pm$	1.421	&	8.88	$\pm$	2.16	&	-53.9	$\pm$	6.7	&	f	\\
138	&	16:26:24.6	&	-24:23:11.10	&	87.122	$\pm$	2.106	&	9.494	$\pm$	1.307	&	-1.679	$\pm$	1.312	&	10.96	$\pm$	1.52	&	-5.0	$\pm$	3.9	&	e	\\
139	&	16:26:30.8	&	-24:23:11.10	&	99.378	$\pm$	2.120	&	7.089	$\pm$	1.338	&	-2.624	$\pm$	1.359	&	7.49	$\pm$	1.36	&	-10.2	$\pm$	5.1	&	g	\\
140	&	16:26:25.5	&	-24:23:11.10	&	320.829	$\pm$	2.114	&	12.169	$\pm$	1.306	&	1.902	$\pm$	1.318	&	3.82	$\pm$	0.41	&	4.4	$\pm$	3.1	&	e	\\
141	&	16:26:29.9	&	-24:23:11.10	&	335.974	$\pm$	2.101	&	14.871	$\pm$	1.338	&	-7.919	$\pm$	1.346	&	5.00	$\pm$	0.40	&	-14.0	$\pm$	2.3	&	g	\\
142	&	16:26:29.0	&	-24:23:11.10	&	686.497	$\pm$	2.134	&	17.110	$\pm$	1.316	&	6.580	$\pm$	1.322	&	2.66	$\pm$	0.19	&	10.5	$\pm$	2.1	&	e	\\
143	&	16:26:26.4	&	-24:23:11.10	&	815.059	$\pm$	2.143	&	14.348	$\pm$	1.306	&	7.082	$\pm$	1.316	&	1.96	$\pm$	0.16	&	13.1	$\pm$	2.4	&	e	\\
144	&	16:26:28.1	&	-24:23:11.10	&	1001.860	$\pm$	2.132	&	20.516	$\pm$	1.311	&	14.985	$\pm$	1.322	&	2.53	$\pm$	0.13	&	18.1	$\pm$	1.5	&	e	\\
145	&	16:26:27.3	&	-24:23:11.10	&	1208.520	$\pm$	2.179	&	16.976	$\pm$	1.303	&	20.285	$\pm$	1.320	&	2.19	$\pm$	0.11	&	25.0	$\pm$	1.4	&	e	\\
146	&	16:26:23.7	&	-24:23:11.09	&	36.854	$\pm$	2.139	&	7.226	$\pm$	1.329	&	-0.484	$\pm$	1.343	&	19.32	$\pm$	3.78	&	-1.9	$\pm$	5.3	&	e	\\
147	&	16:26:22.9	&	-24:23:11.09	&	101.073	$\pm$	2.259	&	8.288	$\pm$	1.362	&	-2.030	$\pm$	1.374	&	8.33	$\pm$	1.36	&	-6.9	$\pm$	4.6	&	f	\\
148	&	16:26:21.1	&	-24:23:11.09	&	110.047	$\pm$	2.299	&	6.195	$\pm$	1.405	&	-5.566	$\pm$	1.409	&	7.46	$\pm$	1.29	&	-21.0	$\pm$	4.8	&	f	\\
149	&	16:26:22.0	&	-24:23:11.09	&	193.122	$\pm$	2.272	&	4.598	$\pm$	1.386	&	-1.217	$\pm$	1.401	&	2.36	$\pm$	0.72	&	-7.4	$\pm$	8.4	&	f	\\
150	&	16:26:20.2	&	-24:23:11.08	&	41.152	$\pm$	2.353	&	2.214	$\pm$	1.423	&	-5.069	$\pm$	1.439	&	12.98	$\pm$	3.57	&	-33.2	$\pm$	7.4	&	f	\\
151	&	16:26:31.6	&	-24:22:59.10	&	37.752	$\pm$	2.203	&	3.501	$\pm$	1.386	&	4.052	$\pm$	1.401	&	13.70	$\pm$	3.79	&	24.6	$\pm$	7.4	&	g	\\
152	&	16:26:24.6	&	-24:22:59.10	&	71.414	$\pm$	2.169	&	5.972	$\pm$	1.330	&	2.543	$\pm$	1.344	&	8.89	$\pm$	1.89	&	11.5	$\pm$	5.9	&	e	\\
153	&	16:26:30.8	&	-24:22:59.10	&	138.114	$\pm$	2.176	&	0.977	$\pm$	1.363	&	-5.578	$\pm$	1.384	&	3.98	$\pm$	1.00	&	-40.0	$\pm$	6.9	&	g	\\
154	&	16:26:25.5	&	-24:22:59.10	&	304.060	$\pm$	2.147	&	13.581	$\pm$	1.329	&	3.681	$\pm$	1.338	&	4.61	$\pm$	0.44	&	7.6	$\pm$	2.7	&	e	\\
155	&	16:26:29.9	&	-24:22:59.10	&	369.449	$\pm$	2.144	&	7.799	$\pm$	1.361	&	-11.722	$\pm$	1.355	&	3.79	$\pm$	0.37	&	-28.2	$\pm$	2.8	&	g	\\
156	&	16:26:26.4	&	-24:22:59.10	&	532.961	$\pm$	2.154	&	20.744	$\pm$	1.328	&	14.188	$\pm$	1.338	&	4.71	$\pm$	0.25	&	17.2	$\pm$	1.5	&	e	\\
157	&	16:26:29.0	&	-24:22:59.10	&	720.738	$\pm$	2.125	&	16.729	$\pm$	1.329	&	4.677	$\pm$	1.349	&	2.40	$\pm$	0.18	&	7.8	$\pm$	2.2	&	e	\\
158	&	16:26:27.3	&	-24:22:59.10	&	744.220	$\pm$	2.160	&	25.034	$\pm$	1.322	&	24.150	$\pm$	1.348	&	4.67	$\pm$	0.18	&	22.0	$\pm$	1.1	&	e	\\
159	&	16:26:28.1	&	-24:22:59.10	&	904.956	$\pm$	2.147	&	26.453	$\pm$	1.329	&	14.877	$\pm$	1.338	&	3.35	$\pm$	0.15	&	14.7	$\pm$	1.3	&	e	\\
160	&	16:26:22.9	&	-24:22:59.09	&	91.979	$\pm$	2.279	&	9.092	$\pm$	1.376	&	-3.091	$\pm$	1.397	&	10.33	$\pm$	1.52	&	-9.4	$\pm$	4.2	&	f	\\
161	&	16:26:21.1	&	-24:22:59.09	&	222.216	$\pm$	2.338	&	13.295	$\pm$	1.406	&	-6.026	$\pm$	1.429	&	6.54	$\pm$	0.64	&	-12.2	$\pm$	2.8	&	f	\\
162	&	16:26:22.0	&	-24:22:59.09	&	345.663	$\pm$	2.325	&	15.068	$\pm$	1.400	&	-3.957	$\pm$	1.413	&	4.49	$\pm$	0.41	&	-7.4	$\pm$	2.6	&	f	\\
163	&	16:26:20.2	&	-24:22:59.08	&	60.825	$\pm$	2.358	&	9.465	$\pm$	1.441	&	-2.125	$\pm$	1.467	&	15.77	$\pm$	2.45	&	-6.3	$\pm$	4.3	&	f	\\
164	&	16:26:45.7	&	-24:22:59.05	&	88.311	$\pm$	3.019	&	4.615	$\pm$	2.132	&	2.294	$\pm$	2.148	&	5.31	$\pm$	2.43	&	13.2	$\pm$	11.9	&	$\cdots$	\\
165	&	16:26:11.4	&	-24:22:59.04	&	48.636	$\pm$	3.084	&	-0.765	$\pm$	1.829	&	-5.855	$\pm$	1.843	&	11.53	$\pm$	3.87	&	-48.7	$\pm$	8.9	&	$\cdots$	\\
166	&	16:26:24.6	&	-24:22:47.10	&	53.074	$\pm$	2.265	&	9.215	$\pm$	1.363	&	1.841	$\pm$	1.366	&	17.52	$\pm$	2.68	&	5.6	$\pm$	4.2	&	e	\\
167	&	16:26:30.8	&	-24:22:47.10	&	134.127	$\pm$	2.222	&	3.846	$\pm$	1.393	&	-9.789	$\pm$	1.399	&	7.77	$\pm$	1.05	&	-34.3	$\pm$	3.8	&	b	\\
168	&	16:26:25.5	&	-24:22:47.10	&	180.227	$\pm$	2.233	&	15.879	$\pm$	1.353	&	10.538	$\pm$	1.373	&	10.55	$\pm$	0.77	&	16.8	$\pm$	2.1	&	e	\\
169	&	16:26:26.4	&	-24:22:47.10	&	381.385	$\pm$	2.224	&	19.126	$\pm$	1.354	&	14.805	$\pm$	1.361	&	6.33	$\pm$	0.36	&	18.9	$\pm$	1.6	&	e	\\
170	&	16:26:29.9	&	-24:22:47.10	&	407.585	$\pm$	2.203	&	6.420	$\pm$	1.378	&	-11.779	$\pm$	1.384	&	3.27	$\pm$	0.34	&	-30.7	$\pm$	2.9	&	b	\\
171	&	16:26:27.3	&	-24:22:47.10	&	585.774	$\pm$	2.234	&	13.418	$\pm$	1.348	&	13.767	$\pm$	1.370	&	3.27	$\pm$	0.23	&	22.9	$\pm$	2.0	&	e	\\
172	&	16:26:29.0	&	-24:22:47.10	&	658.788	$\pm$	2.224	&	7.521	$\pm$	1.370	&	-4.418	$\pm$	1.374	&	1.31	$\pm$	0.21	&	-15.2	$\pm$	4.5	&	e	\\
173	&	16:26:28.1	&	-24:22:47.10	&	739.262	$\pm$	2.188	&	9.876	$\pm$	1.351	&	8.635	$\pm$	1.360	&	1.77	$\pm$	0.18	&	20.6	$\pm$	3.0	&	e	\\
174	&	16:26:23.7	&	-24:22:47.09	&	38.876	$\pm$	2.281	&	8.239	$\pm$	1.371	&	-3.363	$\pm$	1.383	&	22.62	$\pm$	3.78	&	-11.1	$\pm$	4.4	&	e	\\
175	&	16:26:22.9	&	-24:22:47.09	&	159.157	$\pm$	2.341	&	9.114	$\pm$	1.389	&	-4.033	$\pm$	1.406	&	6.20	$\pm$	0.88	&	-11.9	$\pm$	4.0	&	f	\\
176	&	16:26:21.1	&	-24:22:47.09	&	314.588	$\pm$	2.401	&	12.322	$\pm$	1.422	&	-3.301	$\pm$	1.436	&	4.03	$\pm$	0.45	&	-7.5	$\pm$	3.2	&	f	\\
177	&	16:26:22.0	&	-24:22:47.09	&	413.865	$\pm$	2.349	&	8.794	$\pm$	1.403	&	-10.808	$\pm$	1.419	&	3.35	$\pm$	0.34	&	-25.4	$\pm$	2.9	&	f	\\
178	&	16:26:19.3	&	-24:22:47.08	&	59.614	$\pm$	2.493	&	7.615	$\pm$	1.482	&	-4.514	$\pm$	1.498	&	14.64	$\pm$	2.57	&	-15.3	$\pm$	4.8	&	$\cdots$	\\
179	&	16:26:20.2	&	-24:22:47.08	&	156.319	$\pm$	2.393	&	9.025	$\pm$	1.453	&	-2.469	$\pm$	1.475	&	5.91	$\pm$	0.94	&	-7.7	$\pm$	4.5	&	f	\\
180	&	16:26:24.6	&	-24:22:35.10	&	127.395	$\pm$	2.270	&	14.736	$\pm$	1.376	&	2.503	$\pm$	1.397	&	11.68	$\pm$	1.10	&	4.8	$\pm$	2.7	&	e	\\
181	&	16:26:30.8	&	-24:22:35.10	&	212.186	$\pm$	2.256	&	3.442	$\pm$	1.415	&	-19.852	$\pm$	1.426	&	9.47	$\pm$	0.68	&	-40.1	$\pm$	2.0	&	b	\\
182	&	16:26:25.5	&	-24:22:35.10	&	251.386	$\pm$	2.291	&	16.480	$\pm$	1.379	&	8.631	$\pm$	1.404	&	7.38	$\pm$	0.55	&	13.8	$\pm$	2.2	&	e	\\
183	&	16:26:26.4	&	-24:22:35.10	&	413.479	$\pm$	2.280	&	11.576	$\pm$	1.375	&	5.455	$\pm$	1.392	&	3.08	$\pm$	0.33	&	12.6	$\pm$	3.1	&	e	\\
184	&	16:26:29.9	&	-24:22:35.10	&	456.761	$\pm$	2.252	&	5.149	$\pm$	1.392	&	-9.420	$\pm$	1.406	&	2.33	$\pm$	0.31	&	-30.7	$\pm$	3.7	&	b	\\
185	&	16:26:27.3	&	-24:22:35.10	&	455.718	$\pm$	2.279	&	4.230	$\pm$	1.381	&	1.790	$\pm$	1.391	&	0.96	$\pm$	0.30	&	11.5	$\pm$	8.7	&	e	\\
186	&	16:26:23.7	&	-24:22:35.09	&	104.526	$\pm$	2.323	&	11.715	$\pm$	1.390	&	-3.032	$\pm$	1.401	&	11.50	$\pm$	1.35	&	-7.3	$\pm$	3.3	&	e	\\
187	&	16:26:22.9	&	-24:22:35.09	&	178.125	$\pm$	2.360	&	6.971	$\pm$	1.401	&	-1.650	$\pm$	1.422	&	3.94	$\pm$	0.79	&	-6.7	$\pm$	5.7	&	f	\\
188	&	16:26:21.1	&	-24:22:35.09	&	220.112	$\pm$	2.435	&	8.737	$\pm$	1.443	&	-1.179	$\pm$	1.461	&	3.95	$\pm$	0.66	&	-3.8	$\pm$	4.7	&	f	\\
189	&	16:26:22.0	&	-24:22:35.09	&	261.877	$\pm$	2.396	&	8.756	$\pm$	1.426	&	-5.262	$\pm$	1.445	&	3.86	$\pm$	0.55	&	-15.5	$\pm$	4.0	&	f	\\
190	&	16:26:19.3	&	-24:22:35.08	&	56.937	$\pm$	2.484	&	4.730	$\pm$	1.488	&	-2.345	$\pm$	1.503	&	8.90	$\pm$	2.65	&	-13.2	$\pm$	8.1	&	$\cdots$	\\
191	&	16:26:20.2	&	-24:22:35.08	&	125.177	$\pm$	2.428	&	9.520	$\pm$	1.454	&	-1.563	$\pm$	1.467	&	7.62	$\pm$	1.17	&	-4.7	$\pm$	4.4	&	f	\\
192	&	16:26:31.6	&	-24:22:23.10	&	87.998	$\pm$	2.335	&	-0.555	$\pm$	1.462	&	-14.397	$\pm$	1.483	&	16.29	$\pm$	1.74	&	-46.1	$\pm$	2.9	&	b	\\
193	&	16:26:24.6	&	-24:22:23.10	&	235.612	$\pm$	2.314	&	8.555	$\pm$	1.405	&	5.685	$\pm$	1.427	&	4.32	$\pm$	0.60	&	16.8	$\pm$	4.0	&	d	\\
194	&	16:26:30.8	&	-24:22:23.10	&	265.974	$\pm$	2.288	&	1.192	$\pm$	1.444	&	-14.088	$\pm$	1.458	&	5.29	$\pm$	0.55	&	-42.6	$\pm$	2.9	&	b	\\
195	&	16:26:28.1	&	-24:22:23.10	&	321.044	$\pm$	2.315	&	-9.868	$\pm$	1.410	&	-4.661	$\pm$	1.431	&	3.37	$\pm$	0.44	&	-77.4	$\pm$	3.7	&	c	\\
196	&	16:26:25.5	&	-24:22:23.10	&	327.336	$\pm$	2.300	&	7.121	$\pm$	1.404	&	3.980	$\pm$	1.421	&	2.45	$\pm$	0.43	&	14.6	$\pm$	5.0	&	d	\\
197	&	16:26:29.0	&	-24:22:23.10	&	356.454	$\pm$	2.331	&	-5.714	$\pm$	1.432	&	-1.487	$\pm$	1.440	&	1.61	$\pm$	0.40	&	-82.7	$\pm$	7.0	&	c	\\
198	&	16:26:29.9	&	-24:22:23.10	&	376.302	$\pm$	2.288	&	-4.820	$\pm$	1.422	&	-5.187	$\pm$	1.446	&	1.84	$\pm$	0.38	&	-66.4	$\pm$	5.8	&	b	\\
199	&	16:26:27.3	&	-24:22:23.10	&	393.647	$\pm$	2.295	&	-3.922	$\pm$	1.410	&	-4.097	$\pm$	1.413	&	1.40	$\pm$	0.36	&	-66.9	$\pm$	7.1	&	c	\\
200	&	16:26:21.1	&	-24:22:23.09	&	121.082	$\pm$	2.429	&	7.619	$\pm$	1.462	&	3.376	$\pm$	1.476	&	6.78	$\pm$	1.22	&	11.9	$\pm$	5.1	&	d	\\
201	&	16:26:23.7	&	-24:22:23.09	&	163.410	$\pm$	2.324	&	8.735	$\pm$	1.416	&	3.330	$\pm$	1.424	&	5.65	$\pm$	0.87	&	10.4	$\pm$	4.4	&	d	\\
202	&	16:26:22.0	&	-24:22:23.09	&	181.200	$\pm$	2.453	&	3.215	$\pm$	1.440	&	2.708	$\pm$	1.459	&	2.18	$\pm$	0.80	&	20.1	$\pm$	9.9	&	d	\\
203	&	16:26:22.9	&	-24:22:23.09	&	182.764	$\pm$	2.444	&	5.491	$\pm$	1.429	&	6.321	$\pm$	1.438	&	4.51	$\pm$	0.79	&	24.5	$\pm$	4.9	&	d	\\
204	&	16:26:20.2	&	-24:22:23.08	&	52.899	$\pm$	2.470	&	3.222	$\pm$	1.474	&	2.652	$\pm$	1.488	&	7.38	$\pm$	2.82	&	19.7	$\pm$	10.2	&	d	\\
205	&	16:26:32.5	&	-24:22:11.10	&	48.267	$\pm$	2.362	&	-1.336	$\pm$	1.509	&	-9.367	$\pm$	1.509	&	19.35	$\pm$	3.27	&	-49.1	$\pm$	4.6	&	b	\\
206	&	16:26:29.0	&	-24:22:11.10	&	143.786	$\pm$	2.370	&	-10.132	$\pm$	1.461	&	-0.830	$\pm$	1.466	&	7.00	$\pm$	1.02	&	-87.7	$\pm$	4.1	&	c	\\
207	&	16:26:31.6	&	-24:22:11.10	&	154.485	$\pm$	2.387	&	-2.150	$\pm$	1.496	&	-16.003	$\pm$	1.504	&	10.41	$\pm$	0.99	&	-48.8	$\pm$	2.7	&	b	\\
208	&	16:26:28.1	&	-24:22:11.10	&	160.939	$\pm$	2.374	&	-9.426	$\pm$	1.444	&	-2.592	$\pm$	1.456	&	6.01	$\pm$	0.90	&	-82.3	$\pm$	4.3	&	c	\\
209	&	16:26:29.9	&	-24:22:11.10	&	222.641	$\pm$	2.385	&	-3.662	$\pm$	1.455	&	-6.375	$\pm$	1.471	&	3.24	$\pm$	0.66	&	-59.9	$\pm$	5.7	&	b	\\
210	&	16:26:24.6	&	-24:22:11.10	&	234.259	$\pm$	2.338	&	8.571	$\pm$	1.436	&	3.194	$\pm$	1.436	&	3.86	$\pm$	0.61	&	10.2	$\pm$	4.5	&	d	\\
211	&	16:26:30.8	&	-24:22:11.10	&	247.756	$\pm$	2.364	&	-7.141	$\pm$	1.472	&	-18.687	$\pm$	1.488	&	8.05	$\pm$	0.60	&	-55.5	$\pm$	2.1	&	b	\\
212	&	16:26:27.3	&	-24:22:11.10	&	253.879	$\pm$	2.372	&	-10.082	$\pm$	1.438	&	-5.984	$\pm$	1.451	&	4.58	$\pm$	0.57	&	-74.7	$\pm$	3.5	&	c	\\
213	&	16:26:25.5	&	-24:22:11.10	&	268.096	$\pm$	2.378	&	2.588	$\pm$	1.436	&	2.764	$\pm$	1.451	&	1.31	$\pm$	0.54	&	23.4	$\pm$	10.9	&	d	\\
214	&	16:26:22.0	&	-24:22:11.09	&	97.438	$\pm$	2.497	&	0.420	$\pm$	1.441	&	4.959	$\pm$	1.464	&	4.88	$\pm$	1.51	&	42.6	$\pm$	8.3	&	d	\\
215	&	16:26:22.9	&	-24:22:11.09	&	154.229	$\pm$	2.455	&	6.446	$\pm$	1.441	&	9.270	$\pm$	1.460	&	7.26	$\pm$	0.95	&	27.6	$\pm$	3.7	&	d	\\
216	&	16:26:23.7	&	-24:22:11.09	&	186.026	$\pm$	2.417	&	8.161	$\pm$	1.437	&	4.268	$\pm$	1.452	&	4.89	$\pm$	0.78	&	13.8	$\pm$	4.5	&	d	\\
217	&	16:26:28.1	&	-24:21:59.10	&	44.226	$\pm$	2.375	&	-7.979	$\pm$	1.483	&	-0.338	$\pm$	1.496	&	17.74	$\pm$	3.49	&	-88.8	$\pm$	5.4	&	c	\\
218	&	16:26:32.5	&	-24:21:59.10	&	62.207	$\pm$	2.364	&	1.265	$\pm$	1.538	&	-7.122	$\pm$	1.549	&	11.36	$\pm$	2.53	&	-40.0	$\pm$	6.1	&	b	\\
219	&	16:26:29.9	&	-24:21:59.10	&	92.685	$\pm$	2.414	&	-4.357	$\pm$	1.499	&	3.663	$\pm$	1.503	&	5.92	$\pm$	1.63	&	70.0	$\pm$	7.6	&	b	\\
220	&	16:26:27.3	&	-24:21:59.10	&	147.180	$\pm$	2.391	&	-7.698	$\pm$	1.476	&	-8.673	$\pm$	1.491	&	7.81	$\pm$	1.02	&	-65.8	$\pm$	3.7	&	c	\\
221	&	16:26:31.6	&	-24:21:59.10	&	161.920	$\pm$	2.428	&	-1.451	$\pm$	1.507	&	-11.238	$\pm$	1.529	&	6.93	$\pm$	0.95	&	-48.7	$\pm$	3.8	&	b	\\
222	&	16:26:30.8	&	-24:21:59.10	&	190.935	$\pm$	2.390	&	-2.690	$\pm$	1.498	&	-11.650	$\pm$	1.517	&	6.21	$\pm$	0.80	&	-51.5	$\pm$	3.6	&	b	\\
223	&	16:26:26.4	&	-24:21:59.10	&	205.434	$\pm$	2.365	&	-0.648	$\pm$	1.464	&	-5.748	$\pm$	1.466	&	2.72	$\pm$	0.71	&	-48.2	$\pm$	7.3	&	c	\\
224	&	16:26:24.6	&	-24:21:59.10	&	239.257	$\pm$	2.468	&	6.322	$\pm$	1.458	&	1.838	$\pm$	1.478	&	2.68	$\pm$	0.61	&	8.1	$\pm$	6.4	&	d	\\
225	&	16:26:22.9	&	-24:21:59.09	&	148.213	$\pm$	2.557	&	8.257	$\pm$	1.467	&	6.551	$\pm$	1.499	&	7.04	$\pm$	1.01	&	19.2	$\pm$	4.0	&	d	\\
226	&	16:26:23.7	&	-24:21:59.09	&	242.340	$\pm$	2.485	&	10.178	$\pm$	1.458	&	5.633	$\pm$	1.468	&	4.76	$\pm$	0.60	&	14.5	$\pm$	3.6	&	d	\\
227	&	16:26:32.5	&	-24:21:47.10	&	44.850	$\pm$	2.477	&	-5.374	$\pm$	1.581	&	-7.263	$\pm$	1.592	&	19.83	$\pm$	3.71	&	-63.2	$\pm$	5.0	&	b	\\
228	&	16:26:27.3	&	-24:21:47.10	&	49.048	$\pm$	2.464	&	-3.068	$\pm$	1.502	&	-4.613	$\pm$	1.514	&	10.87	$\pm$	3.13	&	-61.8	$\pm$	7.8	&	c	\\
229	&	16:26:30.8	&	-24:21:47.10	&	75.361	$\pm$	2.418	&	-3.993	$\pm$	1.540	&	-5.164	$\pm$	1.555	&	8.41	$\pm$	2.07	&	-63.9	$\pm$	6.8	&	b	\\
230	&	16:26:31.6	&	-24:21:47.10	&	98.617	$\pm$	2.441	&	-6.708	$\pm$	1.555	&	-8.407	$\pm$	1.571	&	10.79	$\pm$	1.61	&	-64.3	$\pm$	4.2	&	b	\\
231	&	16:26:26.4	&	-24:21:47.10	&	141.314	$\pm$	2.503	&	-2.150	$\pm$	1.490	&	-4.842	$\pm$	1.518	&	3.59	$\pm$	1.07	&	-57.0	$\pm$	8.1	&	c	\\
232	&	16:26:22.9	&	-24:21:47.09	&	81.893	$\pm$	2.610	&	2.848	$\pm$	1.503	&	4.478	$\pm$	1.528	&	6.21	$\pm$	1.87	&	28.8	$\pm$	8.2	&	d	\\
233	&	16:26:23.7	&	-24:21:47.09	&	196.629	$\pm$	2.582	&	4.398	$\pm$	1.501	&	2.858	$\pm$	1.505	&	2.56	$\pm$	0.76	&	16.5	$\pm$	8.2	&	d	\\
234	&	16:26:26.4	&	-24:21:35.10	&	84.813	$\pm$	2.582	&	-3.771	$\pm$	1.539	&	-4.027	$\pm$	1.558	&	6.24	$\pm$	1.84	&	-66.6	$\pm$	8.0	&	c	\\
235	&	16:26:23.7	&	-24:21:35.09	&	90.810	$\pm$	2.636	&	3.895	$\pm$	1.538	&	3.342	$\pm$	1.551	&	5.39	$\pm$	1.71	&	20.3	$\pm$	8.6	&	d	\\
236	&	16:26:24.6	&	-24:21:11.10	&	47.959	$\pm$	2.837	&	-4.625	$\pm$	1.611	&	2.377	$\pm$	1.643	&	10.31	$\pm$	3.43	&	76.4	$\pm$	9.0	&	d	\\
237	&	16:26:10.6	&	-24:19:35.03	&	77.061	$\pm$	4.747	&	-2.700	$\pm$	2.625	&	5.578	$\pm$	2.565	&	7.31	$\pm$	3.38	&	57.9	$\pm$	12.1	&	$\cdots$	\\
238	&	16:26:43.9	&	-24:17:35.06	&	124.438	$\pm$	7.065	&	13.784	$\pm$	4.443	&	-2.667	$\pm$	4.528	&	10.70	$\pm$	3.63	&	-5.5	$\pm$	9.2	&	$\cdots$	\\
239	&	16:26:43.1	&	-24:17:23.06	&	134.654	$\pm$	7.249	&	15.535	$\pm$	4.650	&	-0.423	$\pm$	4.745	&	11.01	$\pm$	3.51	&	-0.8	$\pm$	8.7	&	$\cdots$	\\
240	&	16:26:43.9	&	-24:17:23.06	&	170.148	$\pm$	7.541	&	21.291	$\pm$	4.948	&	-2.326	$\pm$	5.059	&	12.25	$\pm$	2.96	&	-3.1	$\pm$	6.8	&	$\cdots$	\\
\enddata
\tablecomments{Units of right ascension are hours, minutes, and seconds,
               and units of declination are degrees, arcminutes,
               and arcseconds.
               Positions are from the Stokes~$I$ image (Figure \ref{fig:OphA_stokesI}), and these sources are sorted by declinations.
Sources with $I > 0$, $P/ \delta P > 2$, and $\delta P < 4$ are listed.
Components refer to regions shown in Figure \ref{fig:region}.}
\end{deluxetable*}
\clearpage

\begin{deluxetable*}{crrc}
\tabletypesize{\small}
\tablewidth{0pt}
\tablecaption{Median $P$, Median $\theta$, and Magnetic Field Direction in Each Component from Figure \ref{fig:region}}
\tablehead{
\colhead{Component}	&  \colhead{$P$}      &  \colhead{ $\theta$ }  		&  \colhead{ $^{\dagger}$MF Direction }   \\
\colhead{           }	&  \colhead{ (\%) }    &  \colhead{ ($\arcdeg$) }  	&  \colhead{ ($\arcdeg$) }   }
\startdata
a 	&	1.44	$\pm$ 0.07	&  144.4  $\pm$ 	1.5	&	54.4		\\
b 	&	6.18	$\pm$ 0.95	&  129.7  $\pm$ 	4.4	&	39.7  	\\
c 	&   	5.19	$\pm$ 0.95	&  106.5  $\pm$ 	5.2	&	16.5  	\\
d 	&   	3.96	$\pm$ 0.80	&   14.0   $\pm$ 5.8	&      104.0 	\\
e 	&   	3.12	$\pm$ 0.29	&    9.6   $\pm$  2.7	&	99.6		\\
f 	&   	4.83	$\pm$ 0.79	&  166.2  $\pm$ 	4.7	&	76.2		\\
g 	&   	8.28	$\pm$ 0.97 	&  156.3  $\pm$ 	3.3	&	66.3		\\
h 	&   	7.92	$\pm$ 2.82 	&  166.9  $\pm$ 	10.2	&	76.9		\\
i	&   	2.76	$\pm$ 0.29	&  165.0  $\pm$ 	3.0	&	75.0		\\
j 	&   	7.96	$\pm$ 2.23 	&   29.9   $\pm$ 8.0	&	119.9		\\
\enddata
\tablecomments{Median polarization degrees $P$ and 
polarization position angles $\theta$ were calculated using Stokes $Q$ and $U$ 
in each component of Figure \ref{fig:region}.
$^{\dagger}$MF: Indicated magnetic field direction, which is $\theta$ rotated by 90\arcdeg.
}
\label{tab:region}
\end{deluxetable*}

\end{document}